\documentclass[12pt,english,aps,manuscript,superscriptaddress]{revtex4}
\usepackage[T1]{fontenc}
\usepackage[latin9]{inputenc}
\usepackage{color}
\usepackage{babel}
\usepackage{verbatim}
\usepackage{amsbsy}
\usepackage{amstext}
\usepackage{graphicx}
\usepackage{esint}
\usepackage[unicode=true,pdfusetitle,
 bookmarks=true,bookmarksnumbered=false,bookmarksopen=false,
 breaklinks=false,pdfborder={0 0 0},pdfborderstyle={},backref=false,colorlinks=true]
 {hyperref}
\hypersetup{
 linkcolor=blue, citecolor=cyan}

\makeatletter

\providecommand{\tabularnewline}{\\}

\@ifundefined{textcolor}{}
{%
 \definecolor{BLACK}{gray}{0}
 \definecolor{WHITE}{gray}{1}
 \definecolor{RED}{rgb}{1,0,0}
 \definecolor{GREEN}{rgb}{0,1,0}
 \definecolor{BLUE}{rgb}{0,0,1}
 \definecolor{CYAN}{cmyk}{1,0,0,0}
 \definecolor{MAGENTA}{cmyk}{0,1,0,0}
 \definecolor{YELLOW}{cmyk}{0,0,1,0}
}

\usepackage{xcolor}

\makeatother

\begin{document}
\begin{flushright}
USTC-ICTS/PCFT-21-25
\par\end{flushright}
\title{Lepton pair production in ultra-peripheral collisions}
\author{Ren-jie Wang}
\email{wrjn@mail.ustc.edu.cn}

\affiliation{Department of Modern Physics, University of Science and Technology
of China, Anhui 230026, China}
\author{Shi Pu}
\email{shipu@ustc.edu.cn}

\affiliation{Department of Modern Physics, University of Science and Technology
of China, Anhui 230026, China}
\author{Qun Wang}
\email{qunwang@ustc.edu.cn}

\affiliation{Interdisciplinary Center for Theoretical Study and Department of Modern
Physics, University of Science and Technology of China, Hefei, Anhui
230026, China}
\affiliation{Peng Huanwu Center for Fundamental Theory, Hefei, Anhui 230026, China}
\begin{abstract}
The lepton pair production in ultra-peripheral collisions (UPC) is
studied in the classical field approximation. We derive a general
form of the cross section in terms of photon distributions which depend
on the transverse momentum and coordinate based on the wave packet
form of nuclear wave functions. Such a general form of the cross section
in the classical field approximation contains the results of the generalized
equivalent photon approximation (EPA) as well as the corrections beyond
EPA in the Born approximation. By rewriting the general form of the
cross section in light-cone coordinates, we find a good connection
with the transverse momentum dependent distribution (TMD) factorization
formalism in the Born approximation. Our numerical results are consistent
with current experimental data.

\end{abstract}
\maketitle

\section{Introduction \label{sec:Introduction}}


Strong electromagnetic fields are produced in the relativistic heavy
ion collisions \citep{Kharzeev:2007jp,Skokov:2009qp,Bzdak:2011yy,Voronyuk:2011jd,Deng:2012pc,Roy:2015coa,Li:2016tel,Inghirami:2016iru,Siddique:2021smf}.
One way to describe the evolution of electromagnetic fields is through
the relativistic magnetohydrodynamics consisting of the hydrodynamical
conservation equations coupled with Maxwell's equations. The solutions
to equations of ideal magnetohydrodynamics with longitudinal boost
invariance show that the transverse magnetic field decays as $\sim1/\tau$
with $\tau$ being the proper time \citep{Pu:2016ayh,Roy:2015kma,Pu:2016rdq,Shokri:2018qcu,Siddique:2019gqh,Wang:2020qpx}
(for a recent numerical simulation of electromagnetic fields based
on ideal magnetohydrodynamics, see, e.g., Ref. \citep{Inghirami:2016iru}).

Although strong electromagnetic fields decay rapidly, there are still
many novel transport phenomena induced by strong fields that could
be measured in experiments, such as the chiral magnetic and separation
effects \citep{Vilenkin:1980fu,Kharzeev:2007jp,Fukushima:2008xe},
the chiral electric separation effect \citep{Huang:2013iia,Pu:2014cwa,Jiang:2014ura,Pu:2014fva},
and other nonlinear chiral transport phenomena \citep{Chen:2016xtg,Pu:2014fva,Hidaka:2017auj,Ebihara:2017suq,Chen:2013dca}.
These effects can be described by chiral kinetic theory for massless
fermions derived from the path integral \citep{Stephanov:2012ki,Chen:2013iga,Chen:2014cla},
the Hamiltonian approaches \citep{Son:2012wh,Son:2012zy}, the quantum
kinetic theory via Wigner functions \citep{Gao:2012ix,Chen:2012ca,Gao:2015zka,Hidaka:2016yjf,Hidaka:2017auj,Gao:2017gfq,Hidaka:2018mel,Gao:2018wmr,Huang:2018wdl},
and the world-line formalism \citep{Mueller:2017lzw}. The chiral
kinetic theory has been extended to the massive fermions and with
collision kernels \citep{Gao:2019znl,Weickgenannt:2019dks,Weickgenannt:2020aaf,Weickgenannt:2020sit,Hattori:2019ahi,Yang:2020hri,Liu:2020flb,Weickgenannt:2021cuo,Sheng:2021kfc}.
For reviews of recent developments in this field, see, e.g., Refs.
\citep{Liao:2014ava,Kharzeev:2015kna,Huang:2015oca,Gao:2020vbh,Liu:2020ymh}.

Another type of effects is related to the non-perturbative production
of lepton pairs in strong electric fields through Schwinger mechanism
\citep{Schwinger:1951nm}. Recent developments along this line include
the lepton pair production in strong magnetic fields by Schwinger
mechanism \citep{Fukushima:2010vw,Warringa:2012bq,Copinger:2018ftr,Copinger:2020nyx}
and the vacuum birefringence \citep{Hattori:2012je,Hattori:2012ny,Hattori:2020htm}.


Recently the lepton pair production through strong electromagnetic
fields in ultra-peripheral collisions (UPC) has drawn broad interest.
Back to 1930s, Weizsacker and Williams considered the electromagnetic
field produced by a fast moving particle as an equivalent flux of
quasi-real photons \citep{vonWeizsacker:1934nji,Williams:1934ad}.
This approximation is called Weizsacker-Williams method or equivalent
photon approximation (EPA) \citep{Jackson:1998nia}. A related process
is the lepton pair production through collisions of two real photons
and was studied by Breit and Wheeler \citep{Breit:1934zz} under the
condition that the total energy of two photons should be greater than
the mass of the lepton pair. The STAR Collaboration at Relativistic
Heavy Ion Collider (RHIC) has measured the lepton pair ($l\overline{l}$)
production process in UPC \citep{Adams:2004rz}. There are also several
measurements related to nonlinear effects of quantum electrodynamics
(QED) such as the vacuum birefringence \citep{Adam:2019mby} and the
light-by-light scattering \citep{Aaboud:2017bwk}. The transverse
momentum spectra of the lepton pair in peripheral collisions of heavy
ions are found to be significantly broader than the ones in UPC by
STAR \citep{Adam:2018tdm} and by the ATLAS collaboration at the Large
Hadron Collider (LHC) \citep{Aaboud:2018eph}. Such broadenings may
arise from medium effects in peripheral collisions, therefore, UPC
may provide a baseline for future studies of medium effects. 


Several theoretical methods are available to describe the lepton pair
production in UPC. A widely used method is the EPA or the generalized
EPA (gEPA). The total cross section of $\gamma\gamma\rightarrow l\overline{l}$
has been calculated in EPA in the classical field approximation of
QED \citep{Vidovic:1992ik}. The original EPA calculations predict
that the position (transverse momentum) of the peak in the transverse
momentum spectrum is less than 20 MeV, inconsistent with the experimental
data \citep{Baltz:2009jk,Klein:2016yzr,Zha:2018ywo}. Therefore the
generalized EPA was proposed \citep{Vidovic:1992ik,Hencken:1994my,Hencken:2004td,Zha:2018tlq,Zha:2018ywo,Brandenburg:2020ozx,Zha:2021jhf}
to give the correct position of the peak in the transverse momentum
spectrum \citep{Zha:2018tlq,Brandenburg:2020ozx,Zha:2021jhf}.

The azimuthal asymmetry in the lepton pair from linearly polarized
photons in UPC has been studied in the TMD factorization formalism
\citep{Li:2019yzy,Li:2019sin} similar to polarized gluons \citep{Metz:2011wb,Akcakaya:2012si,Pisano:2013cya}.
In this formalism, the photon Wigner functions are introduced into
the cross section which depend on the transverse momentum and coordinate
\citep{Klein:2020jom,Xiao:2020ddm}. The broadening of transverse
momentum has also been studied in the TMD formalism \citep{Klein:2018fmp}.


With all these different formulations in different perspectives of
the process, it is natural to ask if there is a unified description.
In this paper, we will derive a general form of the cross section
based on wave-packet nuclear wave functions, which incorporates photon
distributions with the dependence on the transverse momentum and coordinate.
The cross sections in (g)EPA and the TMD formalism can be derived
from the general form. The numerical results of the cross section
in the general form are in a good agreement with experimental data.

This paper is organized as follows. In Sec. \ref{sec:General-expression},
we derive a general form of the cross section in terms of transverse
momentum and coordinate dependent photon distributions. In Sec. \ref{sec:Classical-field-approximation},
we implement the classical field approximation. In Sec. \ref{sec:Connection-to-EPA}
we take the ultra-relativistic limit to reproduce the results of (g)EPA.
In Sec. \ref{sec:TMD-factorization-formalism}, we rewrite the general
form of the cross section in light-cone coordinates and make a connection
with results of the TMD formalism. The numerical results of the cross
section in the general form are given in Sec. \ref{sec:Numerical-results}
and a comparison is made with experimental data as well as the results
of (g)EPA and the TMD formalism. The main results of this work are
summarized in Sec. \ref{sec:Summary}.

Throughout this paper, we choose the metric $g_{\mu\nu}=\textrm{diag}\{+,-,-,-\}$
for ordinary coordinates $(x^{0},x^{1},x^{2},x^{3})=(t,\mathbf{x})$.
The light-cone coordinates $x^{\mu}=(x^{+},x^{-},\mathbf{x}_{T})$
with $x^{\pm}=(x^{0}\pm x^{3})/\sqrt{2}$ and $\mathbf{x}_{T}=(x^{1},x^{2})$
are also used. A vector $a^{\mu}$ can be written as $a^{\mu}=a^{+}n_{+}^{\mu}+a^{-}n_{-}^{\mu}+a_{T}^{\mu}$
where $a^{\pm}=(a^{0}\pm a^{3})/\sqrt{2}$ and $n_{\pm}^{\mu}$ are
light-like vector satisfying $n_{+}^{2}=n_{-}^{2}=0$ and $n_{+}\cdot n_{-}=1$.
The inner product of two vectors in light-cone coordinates is $a\cdot b=a^{-}b^{+}+a^{+}b^{-}-\boldsymbol{a}_{T}\cdot\boldsymbol{b}_{T}$.


\section{General form of cross sections for lepton pairs \label{sec:General-expression}}

In this section, we give a general form for the differential cross
section of lepton pairs in UPC, with the detailed derivation being
given in Appendix \ref{sec:Derive}.

\begin{figure}[t]
\includegraphics[scale=0.45]{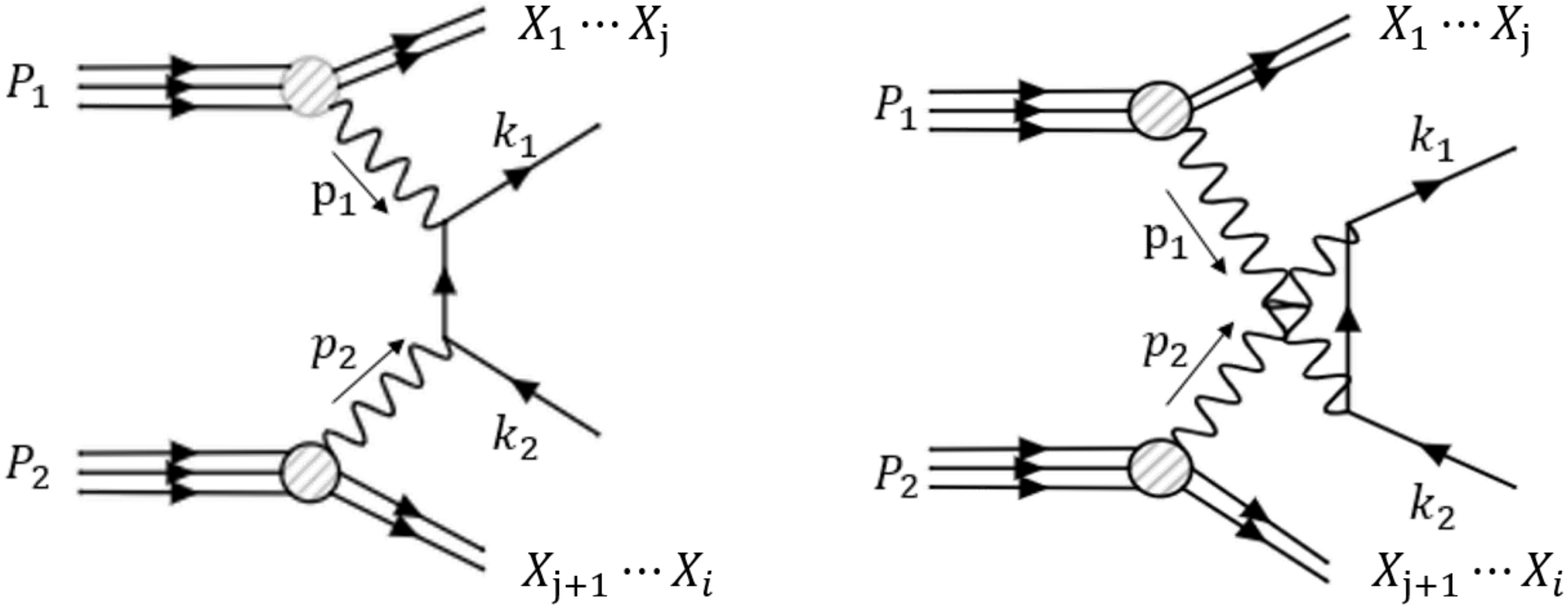}

\caption{Feynman diagrams for the photon fusion process in UPC. \label{fig:Feynman01}}
\end{figure}

\begin{figure}[t]
\includegraphics[scale=0.35]{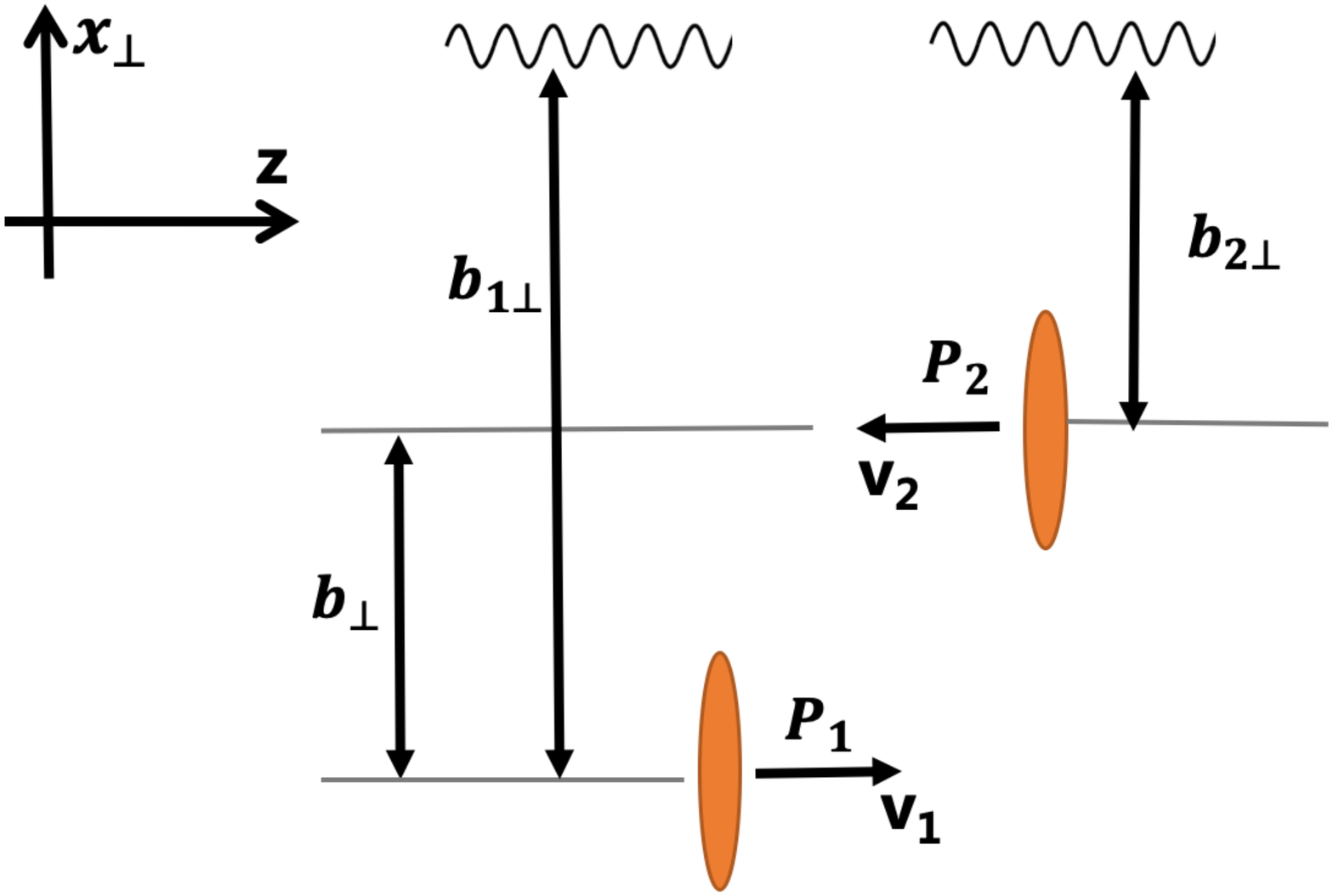}

\caption{A cartoon for photon emission in UPC. \label{fig:collision}}
\end{figure}

As shown in Fig. \ref{fig:collision}, we consider collisions of two
nuclei $A_{1}$ and $A_{2}$ moving alone $\pm z$ direction which
are displaced by an impact parameter $\mathbf{b}_{T}$ and generate
a pair of leptons $l$ and $\overline{l}$ along with other particles
$X_{1},\cdots,X_{f},\cdots$,

\begin{equation}
A_{1}(P_{A1})+A_{2}(P_{A2})\rightarrow l(k_{1})+\overline{l}(k_{2})+\sum_{f}X_{f}(K_{f}),
\end{equation}
where four-momenta of particles are given in parentheses. Here $P_{A1}^{\mu}=(E_{A1},\mathbf{P}_{A1})$
and $P_{A2}^{\mu}=(E_{A2},\mathbf{P}_{A2})$ with $E_{A1}=\sqrt{\mathbf{P}_{A1}^{2}+M_{1}^{2}}$
and $E_{A2}=\sqrt{\mathbf{P}_{A2}^{2}+M_{2}^{2}}$ are on-shell momenta
of two nuclei with masses $M_{1}$ and $M_{2}$ respectively. The
three-momenta of two nuclei are $\mathbf{P}_{A1}=(0,0,P_{A1}^{z})$
and $\mathbf{P}_{A2}=(0,0,-P_{A1}^{z})$ in the center of mass frame
of the collision.

In order to describe collisions at fixed impact parameters, we need
to assume the wave functions of colliding nuclei to be wave packets.
We follow the standard way in quantum field theory to obtain the cross
section at the impact parameter $\mathbf{b}_{T}$ in Eq. (\ref{eq:cross-bt}).
Making the ansatz (\ref{eq:wave_ansatz}) for the longtitudinal momentum
amplitudes of wave-packets and completing the integrals over longitudinal
momenta of wave-packets we arrive at 
\begin{eqnarray}
\sigma & = & \frac{1}{8(2\pi)^{8}}\int d^{2}\mathbf{b}_{T}d^{2}\mathbf{b}_{1T}d^{2}\mathbf{b}_{2T}\sum_{\{f\}}\int\frac{d^{3}k_{1}}{(2\pi)^{3}2E_{k1}}\frac{d^{3}k_{2}}{(2\pi)^{3}2E_{k2}}\prod_{f}\frac{d^{3}K_{f}}{(2\pi)^{3}2E_{Kf}}\nonumber \\
 &  & \times\int d^{2}\mathbf{P}_{1T}d^{2}\mathbf{P}_{2T}d^{2}\mathbf{P}_{1T}^{\prime}d^{2}\mathbf{P}_{2T}^{\prime}\frac{1}{v\sqrt{E_{P1}E_{P2}E_{P1^{\prime}}E_{P2^{\prime}}}}\nonumber \\
 &  & \times G^{2}\left[(P_{1}^{\prime z}-P_{A1}^{z})^{2}\right]\phi_{T}(\mathbf{P}_{1T})\phi_{T}(\mathbf{P}_{2T})\phi_{T}^{*}(\mathbf{P}_{1T}^{\prime})\phi_{T}^{*}(\mathbf{P}_{2T}^{\prime})\nonumber \\
 &  & \times e^{-i\mathbf{b}_{1T}\cdot\boldsymbol{\Delta}_{1T}}e^{-i\mathbf{b}_{2T}\cdot\boldsymbol{\Delta}_{2T}}\delta^{(2)}\left(\mathbf{b}_{T}-\mathbf{b}_{1T}+\mathbf{b}_{2T}\right)\nonumber \\
 &  & \times(2\pi)^{4}\delta^{(4)}\left(P_{1}+P_{2}-k_{1}-k_{2}-\sum_{f}K_{f}\right)\nonumber \\
 &  & \times\sum_{\textrm{spin of }l,\overline{l}}\mathcal{M}_{P_{1}+P_{2}\rightarrow k_{1}+k_{2}+\sum_{f}K_{f}}\mathcal{M}_{P_{1}^{\prime}+P_{2}^{\prime}\rightarrow k_{1}+k_{2}+\sum_{f}K_{f}}^{*}.\label{eq:cross-section-4}
\end{eqnarray}
Here $P_{1}$, $P_{2}$, $P_{1}^{\prime}$ and $P_{2}^{\prime}$ are
on-shell momenta of nuclear wave-packets given by $P_{1}=(E_{P1},\mathbf{P}_{1T},2P_{A1}^{z}-P_{1}^{\prime z})$,
$P_{2}=(E_{P2},\mathbf{P}_{2T},-2P_{A1}^{z}+P_{1}^{\prime z})$, $P_{1}^{\prime}=(E_{P1^{\prime}},\mathbf{P}_{1T}^{\prime},P_{1}^{\prime z})$
and $P_{2}^{\prime}=(E_{P2^{\prime}},\mathbf{P}_{2T}^{\prime},-P_{1}^{\prime z})$.
As the solution to the energy conservation $P_{1}^{\prime z}$ is
a function of transverse momenta $\mathbf{P}_{1T}$, $\mathbf{P}_{2T}$,
$\mathbf{P}_{1T}^{\prime}$ and $\mathbf{P}_{2T}^{\prime}$. We see
that the $z$ components of $P_{1}+P_{2}$ and $P_{1}^{\prime}+P_{2}^{\prime}$
are vanishing. The function $G(x^{2})$ is defined in Eq. (\ref{eq:wave_ansatz}),
which is a positive and decreasing function of $x^{2}$ satisfying
$G(0)=1$. We have used the shifts of transverse momenta $\boldsymbol{\Delta}_{1T}\equiv\mathbf{P}_{1T}^{\prime}-\mathbf{P}_{1T}$
and $\boldsymbol{\Delta}_{2T}\equiv\mathbf{P}_{2T}^{\prime}-\mathbf{P}_{2T}$,
and relative velocity of two nuclei $v$ given in Eq. (\ref{eq:p1z-p2z-prime})
as a function of transverse momenta.

In the tree level of Feynman diagrams (sometimes called Born approximation)
as shown in Fig. \ref{fig:Feynman01}, the lepton pair is produced
in the photon fusion process 
\begin{equation}
\gamma(p_{1})+\gamma(p_{2})\rightarrow l(k_{1})+\overline{l}(k_{2}).
\end{equation}
Here, we assume the photon $\gamma(p_{1})$ or $\gamma(p_{2})$ comes
from the nuclei $A_{1}(P_{1})$ or $A_{2}(P_{2})$, respectively.
Note that, each photon does not have to come from the nuclear center.
We can identify $\mathbf{b}_{iT}$ in Eq. (\ref{eq:cross-section-4})
as the transverse distance between $\gamma(p_{i})$ and $A_{i}(P_{i})$
with $i=1,2$, which are related to the impact parameter $\mathbf{b}_{T}=\mathbf{b}_{1T}-\mathbf{b}_{2T}$
as shown in Fig. \ref{fig:collision}.

The invariant amplitude $\mathcal{M}$ can be obtained through the
matrix element of the operator $\widehat{T}$ or the T-matrix element.
The results are given in Eqs. (\ref{eq:it-matrix},\ref{eq:it-star-matrix}).
By inserting these results for invariant amplitudes into Eq. (\ref{eq:cross-section-4}),
the differential cross section can be put into the form 
\begin{eqnarray}
\frac{d\sigma}{d^{3}k_{1}d^{3}k_{2}} & \approx & \frac{1}{32(2\pi)^{6}}\frac{1}{E_{k1}E_{k2}}\int d^{2}\mathbf{b}_{T}d^{2}\mathbf{b}_{1T}d^{2}\mathbf{b}_{2T}\int d^{4}p_{1}d^{4}p_{2}\nonumber \\
 &  & \times\delta^{(2)}\left(\mathbf{b}_{T}-\mathbf{b}_{1T}+\mathbf{b}_{2T}\right)(2\pi)^{4}\delta^{(4)}\left(p_{1}+p_{2}-k_{1}-k_{2}\right)\nonumber \\
 &  & \times\int\frac{d^{2}\mathbf{P}_{(1+1^{\prime})T}}{(2\pi)^{2}}\frac{d^{2}\mathbf{P}_{(2+2^{\prime})T}}{(2\pi)^{2}}\frac{1}{v\sqrt{E_{P1}E_{P2}E_{P1^{\prime}}E_{P2^{\prime}}}}\nonumber \\
 &  & \times G^{2}\left[(P_{1}^{\prime z}-P_{A1}^{z})^{2}\right]\phi_{T}(\mathbf{P}_{1T})\phi_{T}(\mathbf{P}_{2T})\phi_{T}^{*}(\mathbf{P}_{1T}^{\prime})\phi_{T}^{*}(\mathbf{P}_{2T}^{\prime})\nonumber \\
 &  & \times\mathcal{S}_{\sigma\mu}(p_{1},\mathbf{b}_{1T})\mathcal{S}_{\rho\nu}(p_{2},\mathbf{b}_{2T})\nonumber \\
 &  & \times L^{\mu\nu;\sigma\rho}\left(p_{1},p_{2};p_{1}-P_{1}+P_{1}^{\prime},p_{2}-P_{2}+P_{2}^{\prime};k_{1},k_{2}\right),\label{eq:diff_02}
\end{eqnarray}
where $P_{1}^{\prime z}$ is a function of nulcear transverse momenta
which approaches $P_{A1}^{z}$ if all nulcear transverse momenta are
vanishing, $L^{\mu\nu;\sigma\rho}$ is the lepton tensor in Eq. (\ref{eq:l-mu-nu-1}),
and we have used variables $\boldsymbol{\Delta}_{1T}$ and $\boldsymbol{\Delta}_{2T}$
with 
\begin{eqnarray}
\mathbf{P}_{(1+1^{\prime})T} & = & \frac{1}{2}(\mathbf{P}_{1T}+\mathbf{P}_{1T}^{\prime}),\nonumber \\
\mathbf{P}_{(2+2^{\prime})T} & = & \frac{1}{2}(\mathbf{P}_{2T}+\mathbf{P}_{2T}^{\prime}),
\end{eqnarray}
to replace $\mathbf{P}_{1T}$, $\mathbf{P}_{2T}$, $\mathbf{P}_{1T}^{\prime}$
and $\mathbf{P}_{2T}^{\prime}$. We have used the Wigner functions
for photons 
\begin{eqnarray}
\mathcal{S}_{\sigma\mu}(p_{1},\mathbf{b}_{1T}) & \equiv & \int\frac{d^{2}\boldsymbol{\Delta}_{1T}}{(2\pi)^{2}}\int\frac{d^{4}y_{1}}{(2\pi)^{4}}e^{ip_{1}\cdot y_{1}}\left\langle P_{1}^{\prime}\right|A_{\sigma}^{\dagger}(0)A_{\mu}\left(y_{1}\right)\left|P_{1}\right\rangle e^{-i\mathbf{b}_{1T}\cdot\boldsymbol{\Delta}_{1T}},\nonumber \\
\mathcal{S}_{\rho\nu}(p_{2},\mathbf{b}_{2T}) & \equiv & \int\frac{d^{2}\boldsymbol{\Delta}_{2T}}{(2\pi)^{2}}\int\frac{d^{4}y_{2}}{(2\pi)^{4}}e^{ip_{2}\cdot y_{2}}\left\langle P_{2}^{\prime}\right|A_{\rho}^{\dagger}\left(0\right)A_{\nu}\left(y_{2}\right)\left|P_{2}\right\rangle e^{-i\mathbf{b}_{2T}\cdot\boldsymbol{\Delta}_{2T}},\label{eq:S_p_b_01}
\end{eqnarray}
which are similar to those in Ref. \citep{Klein:2020jom,Xiao:2020ddm}.
We will discuss them carefully in the TMD factorization formalism
in Sec. \ref{sec:TMD-factorization-formalism}.

We can also integrate Eq. (\ref{eq:diff_02}) over $\mathbf{b}_{1T}$
and $\mathbf{b}_{2T}$ and express the differential cross section
at fixed $\mathbf{b}_{T}$ 
\begin{eqnarray}
\frac{d\sigma}{d^{3}k_{1}d^{3}k_{2}d^{2}\mathbf{b}_{T}} & = & \frac{1}{32(2\pi)^{6}}\frac{1}{E_{k1}E_{k2}}\int d^{4}p_{1}d^{4}p_{2}\nonumber \\
 &  & \times\int\frac{d^{2}\mathbf{P}_{1T}}{(2\pi)^{2}}\frac{d^{2}\mathbf{P}_{2T}}{(2\pi)^{2}}\frac{d^{2}\mathbf{P}_{1T}^{\prime}}{(2\pi)^{2}}\frac{d^{2}\mathbf{P}_{2T}^{\prime}}{(2\pi)^{2}}\frac{1}{v\sqrt{E_{P1}E_{P2}E_{P1^{\prime}}E_{P2^{\prime}}}}\nonumber \\
 &  & \times e^{-i\mathbf{b}_{T}\cdot\boldsymbol{\Delta}_{1T}}G^{2}\left[(P_{1}^{\prime z}-P_{A1}^{z})^{2}\right]\phi_{T}(\mathbf{P}_{1T})\phi_{T}(\mathbf{P}_{2T})\phi_{T}^{*}(\mathbf{P}_{1T}^{\prime})\phi_{T}^{*}(\mathbf{P}_{2T}^{\prime})\nonumber \\
 &  & \times L^{\mu\nu;\sigma\rho}\left(p_{1},p_{2};p_{1}-P_{1}+P_{1}^{\prime},p_{2}-P_{2}+P_{2}^{\prime};k_{1},k_{2}\right)\nonumber \\
 &  & \times\int\frac{d^{4}y_{1}}{(2\pi)^{4}}e^{ip_{1}\cdot y_{1}}\left\langle P_{1}^{\prime}\right|A_{\sigma}^{\dagger}(0)A_{\mu}\left(y_{1}\right)\left|P_{1}\right\rangle \nonumber \\
 &  & \times\int\frac{d^{4}y_{2}}{(2\pi)^{4}}e^{ip_{2}\cdot y_{2}}\left\langle P_{2}^{\prime}\right|A_{\rho}^{\dagger}\left(0\right)A_{\nu}\left(y_{2}\right)\left|P_{2}\right\rangle \nonumber \\
 &  & \times(2\pi)^{2}\delta^{(2)}\left(\mathbf{P}_{1T}+\mathbf{P}_{2T}-\mathbf{P}_{1T}^{\prime}-\mathbf{P}_{2T}^{\prime}\right)\nonumber \\
 &  & \times(2\pi)^{4}\delta^{(4)}(p_{1}+p_{2}-k_{1}-k_{2}),\label{eq:diff_04}
\end{eqnarray}
We can also integrate Eq. (\ref{eq:diff_04}) over $\mathbf{b}_{T}$
to obtain
\begin{eqnarray}
\frac{d\sigma}{d^{3}k_{1}d^{3}k_{2}} & = & \frac{1}{32(2\pi)^{6}}\frac{1}{E_{k1}E_{k2}}\int d^{4}p_{1}d^{4}p_{2}\nonumber \\
 &  & \times\int\frac{d^{2}\mathbf{P}_{1T}}{(2\pi)^{2}}\frac{d^{2}\mathbf{P}_{2T}}{(2\pi)^{2}}\frac{1}{vE_{P1}E_{P2}}\left|\phi_{T}(\mathbf{P}_{1T})\right|^{2}\left|\phi_{T}(\mathbf{P}_{2T})\right|^{2}\nonumber \\
 &  & \times S_{\sigma\mu}(P_{1},p_{1})S_{\rho\nu}(P_{2},p_{2})L^{\mu\nu;\sigma\rho}\left(p_{1},p_{2};p_{1},p_{2};k_{1},k_{2}\right)\nonumber \\
 &  & \times(2\pi)^{4}\delta^{4}(p_{1}+p_{2}-k_{1}-k_{2}),\label{eq:diff_05}
\end{eqnarray}
where we have used $P_{1}^{\prime z}=P_{A1}^{z}$ as the solution
to the energy conservation equation with the conditions $\mathbf{P}_{1T}=\mathbf{P}_{1T}^{\prime}$
and $\mathbf{P}_{2T}=\mathbf{P}_{2T}^{\prime}$, so the on-shell momenta
of nuclear wave-packets now become $P_{1}=(E_{P1},\mathbf{P}_{1T},P_{A1}^{z})$
and $P_{2}=(E_{P2},\mathbf{P}_{2T},-P_{A1}^{z})$. We have also defined
\begin{equation}
S_{\sigma\mu}(P,p)\equiv\int\frac{d^{4}y}{(2\pi)^{4}}e^{ip\cdot y}\left\langle P\right|A_{\sigma}^{\dagger}(0)A_{\mu}(y)\left|P\right\rangle ,\label{eq:TMD_unpol_01}
\end{equation}
as the TMD correlation function for unpolarized nuclei.

In this section, we have given a general form for the impact parameter
dependent cross section. The cross sections in Eqs. (\ref{eq:diff_04},
\ref{eq:diff_05}) are similar to the TMD factorization for photons.
We see that $S_{\sigma\mu}(P,p)$ appear in Eqs. (\ref{eq:diff_04},\ref{eq:diff_05})
as non-perturbative soft correlation functions for photons. In the
next section, we will show how to handle soft correlation functions
in the classical field approximation.


\section{Classical field approximation \label{sec:Classical-field-approximation}}

In this section, we take the classical field approximation for the
photon field $A^{\mu}(x)$. We follow the idea in early works \citep{Vidovic:1992ik,Hencken:1994my,Hencken:2004td}
and extend their original formalism by including impact parameters
$\mathbf{b}_{1T}$ and $\mathbf{b}_{2T}$.

We consider collisions of two nuclei in the center of mass frame as
shown in Fig. \ref{fig:collision}. The fluid velocities of two nuclei
are $u_{1}^{\mu}=\gamma_{1}\left(1,0,0,v_{1}\right)$ and $u_{2}^{\mu}=\gamma_{2}\left(1,0,0,-v_{2}\right)$
with $\gamma_{1,2}=1/\sqrt{1-v_{1,2}^{2}}$ being the Lorentz factor.
At very high energy, these four-velocities have properties $u_{1}^{\mu}\propto n_{+}^{\mu}$
and $u_{2}^{\mu}\propto n_{-}^{\mu}$, where $n_{+}^{\mu}$ and $n_{-}^{\mu}$
are light-like vectors given by Eq. (\ref{eq:sudakov-vector}). The
charge currents are then $j_{1,2}^{\mu}=Z_{1,2}e\rho_{1,2}(\mathbf{x})u_{1,2}^{\mu}$,
where $\rho_{1,2}(\mathbf{x})$ are the charge densities. By choosing
the Lorentz gauge $\partial_{\mu}A^{\mu}=0$ or $p\cdot A(p)=0$,
we can solve the Maxwell equations $\partial_{\rho}\partial^{\rho}A_{(1,2)}^{\mu}=j_{1,2}^{\mu}$
to derive the classical photon (electromagnetic or EM) fields $A_{(1,2)}^{\mu}$.
We introduce the nuclear form factor $F(\mathbf{k})$ as the Fourier
transformation of $\rho(\mathbf{x})$ 
\begin{eqnarray}
F(\mathbf{p}) & = & \int d^{3}xe^{-i\mathbf{p}\cdot\boldsymbol{x}}\rho(\mathbf{x}),\nonumber \\
\rho(\mathbf{x}) & = & \int\frac{d^{3}p}{(2\pi)^{3}}e^{i\mathbf{p}\cdot\mathbf{x}}F(\mathbf{p}),
\end{eqnarray}
Then the classical photon fields in momentum space are 
\begin{equation}
A_{(1,2)}^{\mu}(p)=2\pi Z_{1,2}e\delta(p\cdot u_{1,2})\frac{F(p)}{-p^{2}}u_{1,2}^{\mu},\label{eq:classical_01}
\end{equation}
where we have replaced $\delta(p^{0})$ by its covariant form $\delta(p\cdot u_{1,2})$
\citep{Krauss:1997vr}. Note that in Refs. \citep{Vidovic:1992ik,Hencken:1994my},
an extra phase factor $e^{i\mathbf{p}_{T}\cdot\mathbf{b}_{T}}$ is
included into $A_{(1)}^{\mu}(p)$ to describe the dependence on the
impact parameter. Since we have already introduced the dependence
on impact parameters systematically into the differential cross sections
(\ref{eq:diff_02},\ref{eq:diff_04}), we do not need to add this
phase factor into $A_{(1,2)}^{\mu}(p)$ in (\ref{eq:classical_01}).
One can verfiy that our formalism is equivalent to Ref. \citep{Vidovic:1992ik,Hencken:1994my}. 

The matrix elements of photon operators in Eq. (\ref{eq:cross-section-2})
are assumed to take the form with $i=1,2$ labeling two nuclei
\begin{eqnarray}
\left\langle P_{i}^{\prime}\right|A_{\sigma}^{\dagger}(p_{i}^{\prime})A_{\mu}(p_{i})\left|P_{i}\right\rangle  & \approx & A_{\sigma}^{*}(p_{i}^{\prime})A_{\mu}(p_{i})2\sqrt{(P_{i}\cdot u_{i})(P_{i}^{\prime}\cdot u_{i})}\nonumber \\
 &  & \times(2\pi)^{3}\delta^{(3)}\left(\overline{p}_{i}-\overline{p}_{i}^{\prime}-\overline{P}_{i}+\overline{P}_{i}^{\prime}\right)\nonumber \\
 & = & A_{\sigma}^{*}(p_{i}^{\prime})A_{\mu}(p_{i})\frac{1}{\gamma}2\sqrt{E_{Pi}E_{Pi^{\prime}}}\nonumber \\
 &  & \times(2\pi)^{3}\delta\left(p_{i}\cdot\overline{u}_{i}-p_{i}^{\prime}\cdot\overline{u}_{i}-P_{i}\cdot\overline{u}_{i}+P_{i}^{\prime}\cdot\overline{u}_{i}\right)\nonumber \\
 &  & \times\delta^{(2)}\left(\mathbf{p}_{iT}-\mathbf{p}_{iT}^{\prime}-\mathbf{P}_{iT}+\mathbf{P}_{iT}^{\prime}\right),\label{eq:photon-corr}
\end{eqnarray}
where $A_{\sigma}^{*}(p_{i}^{\prime})\equiv A_{\sigma(i)}^{*}(p_{i}^{\prime})$
and $A_{\mu}(p_{i})\equiv A_{\mu(i)}(p_{i})$ denote the classical
photon fields in (\ref{eq:classical_01}), and $\overline{a}_{i}^{\mu}$
denote the components of $a_{i}^{\mu}$ for $a=p,p^{\prime},P,P^{\prime}$
that are perpendicular to $u_{i}^{\mu}$, $\overline{a}_{i}\cdot u_{i}=0$,
which are given by 
\begin{equation}
\overline{a}_{i}^{\mu}\equiv-(a_{i}\cdot\overline{u}_{i})\overline{u}_{i}^{\mu}+g_{T}^{\mu\nu}a_{i,\nu}=\Delta_{u_{i}}^{\mu\nu}a_{i,\nu},
\end{equation}
with $\overline{u}_{i}^{\mu}\equiv\gamma(v,0,0,1)$, $g_{T}^{\mu\nu}\equiv\mathrm{diag}(0,-1,-1,0)$,
and $\Delta_{u_{i}}^{\mu\nu}\equiv g^{\mu\nu}-u_{i}^{\mu}u_{i}^{\nu}$.
Note that $\overline{a}_{i}^{\mu}$ are actually the spatial components
of $a_{i}^{\mu}$ in the comoving frame of the nucleus labeled by
$i$. So $a_{i}^{\mu}$ can be decomposed as $a_{i}^{\mu}=(a_{i}\cdot u_{i})u_{i}^{\mu}+\overline{a}_{i}^{\mu}$.
The explicit forms of $\overline{a}_{i}^{\mu}$ for $a=p,p^{\prime}$
are given by $a_{i}^{\mu}$ under the conditions $a_{i}\cdot u_{i}=0$
\begin{eqnarray}
\overline{a}_{1}^{\mu} & = & \left(a_{1}^{0},\boldsymbol{a}_{1T},\frac{a_{1}^{0}}{v_{1}}\right),\nonumber \\
\overline{a}_{2}^{\mu} & = & \left(a_{2}^{0},\boldsymbol{a}_{2T},-\frac{a_{2}^{0}}{v_{2}}\right).\label{eq:orth-momenta}
\end{eqnarray}

Now we can derive the cross section from Eq. (\ref{eq:cross-section-2})
for collisions of two equal nuclei ($Z_{1}=Z_{2}=Z$ and $v_{1}=v_{2}=v$)
by using Eqs. (\ref{eq:classical_01},\ref{eq:photon-corr}) in Eq.
(\ref{eq:cross-section-2}) and completing the integrals over $p_{1}^{\prime}$,
$p_{2}^{\prime}$, $p_{1}\cdot u_{1}$ and $p_{2}\cdot u_{2}$. The
result is 
\begin{eqnarray}
\sigma & \approx & \frac{e^{4}Z^{4}}{2\gamma^{2}v}\int d^{2}\mathbf{b}_{T}d^{2}\mathbf{b}_{1T}d^{2}\mathbf{b}_{2T}\int\frac{d^{3}k_{1}}{(2\pi)^{3}2E_{k1}}\frac{d^{3}k_{2}}{(2\pi)^{3}2E_{k2}}\int\frac{d^{3}\overline{p}_{1}}{(2\pi)^{3}}\frac{d^{3}\overline{p}_{2}}{(2\pi)^{3}}\nonumber \\
 &  & \times\int\frac{d^{2}\mathbf{P}_{1T}}{(2\pi)^{2}}\frac{d^{2}\mathbf{P}_{2T}}{(2\pi)^{2}}\frac{d^{2}\mathbf{P}_{1T}^{\prime}}{(2\pi)^{2}}\frac{d^{2}\mathbf{P}_{2T}^{\prime}}{(2\pi)^{2}}G^{2}\left[(P_{1}^{\prime z}-P_{A1}^{z})^{2}\right]\nonumber \\
 &  & \times\phi_{T}(\mathbf{P}_{1T})\phi_{T}(\mathbf{P}_{2T})\phi_{T}^{*}(\mathbf{P}_{1T}^{\prime})\phi_{T}^{*}(\mathbf{P}_{2T}^{\prime})\nonumber \\
 &  & \times e^{-i\mathbf{b}_{1T}\cdot\boldsymbol{\Delta}_{1T}}e^{-i\mathbf{b}_{2T}\cdot\boldsymbol{\Delta}_{2T}}\delta^{(2)}\left(\mathbf{b}_{T}-\mathbf{b}_{1T}+\mathbf{b}_{2T}\right)\nonumber \\
 &  & \times\frac{F(-\overline{p}_{1}^{2})}{-\overline{p}_{1}^{2}}\frac{F(-\overline{p}_{2}^{2})}{-\overline{p}_{2}^{2}}\frac{F^{*}(-\overline{p}_{1}^{\prime2})}{-\overline{p}_{1}^{\prime2}}\frac{F^{*}(-\overline{p}_{2}^{\prime2})}{-\overline{p}_{2}^{\prime2}}\nonumber \\
 &  & \times\sum_{\textrm{spin of }l,\overline{l}}\left[u_{1\mu}u_{2\nu}L^{\mu\nu}(\overline{p}_{1},\overline{p}_{2};k_{1},k_{2})\right]\left[u_{1\sigma}u_{2\rho}L^{\sigma\rho*}(\overline{p}_{1}^{\prime},\overline{p}_{2}^{\prime};k_{1},k_{2})\right]\nonumber \\
 &  & \times(2\pi)^{4}\delta^{(4)}\left(\overline{p}_{1}+\overline{p}_{2}-k_{1}-k_{2}\right),\label{eq:cross-section-class-1}
\end{eqnarray}
where $\overline{p}_{1}$, $\overline{p}_{2}$, $\overline{p}_{1}^{\prime}$
and $\overline{p}_{2}^{\prime}$ are given by Eq. (\ref{eq:orth-momenta}),
and $\overline{p}_{i}^{\prime}=\overline{p}_{i}-\overline{P}_{i}+\overline{P}_{i}^{\prime}$
for $i=1,2$. Note that $\overline{p}_{i}$, $\overline{p}_{i}^{\prime}$,
$\overline{P}_{i}$ and $\overline{P}_{i}^{\prime}$ are momenta perpendicular
to the four-velocity $u_{i}^{\mu}$ which have three independent components,
while quantities with index $T$ are transverse ones perpendicular
to the beam direction. We denote the 0-components of photon momenta
$p_{i}$ and $p_{i}^{\prime}$ as $\omega_{i}$ and $\omega_{i}^{\prime}$
for $i=1,2$, respectively, then the conditions $\overline{p}_{i}^{\prime}=\overline{p}_{i}-\overline{P}_{i}+\overline{P}_{i}^{\prime}$
require $\omega_{i}^{\prime}=\omega_{i}-E_{Pi}+E_{Pi^{\prime}}$ and
$\mathbf{p}_{iT}^{\prime}=\mathbf{p}_{iT}-\mathbf{P}_{iT}+\mathbf{P}_{iT}^{\prime}$.
In the derivation of (\ref{eq:cross-section-class-1}), the four-momentum
integrals are treated as $d^{4}p_{i}=d(p_{i}\cdot u_{i})d^{3}\overline{p}_{i}\equiv d(p_{i}\cdot u_{i})d(-p_{i}\cdot\overline{u}_{i})d^{2}\mathbf{p}_{iT}$,
because of $-p_{i}\cdot\overline{u}_{i}=\omega_{i}/(v\gamma)$ when
applying $p_{i}\cdot u_{i}=0$, we have $d^{4}p_{i}=d(p_{i}\cdot u_{i})d\omega_{i}d^{2}\mathbf{p}_{iT}/(v\gamma)$.

Completing the inetgrals over $\mathbf{b}_{1T}$, $\mathbf{b}_{2T}$,
$\mathbf{b}_{T}$, $\mathbf{P}_{iT}$ and $\mathbf{P}_{iT}^{\prime}$
in Eq. (\ref{eq:cross-section-class-1}), we obtain 
\begin{eqnarray}
\sigma & = & \frac{e^{4}Z^{4}}{2\gamma^{4}v^{3}}\int\frac{d^{3}k_{1}}{(2\pi)^{3}2E_{k1}}\frac{d^{3}k_{2}}{(2\pi)^{3}2E_{k2}}\int\frac{d\omega_{1}d^{2}\mathbf{p}_{1T}}{(2\pi)^{3}}\int\frac{d\omega_{2}d^{2}\mathbf{p}_{2T}}{(2\pi)^{3}}\nonumber \\
 &  & \times\frac{F(-\overline{p}_{1}^{2})}{-\overline{p}_{1}^{2}}\frac{F(-\overline{p}_{2}^{2})}{-\overline{p}_{2}^{2}}\frac{F^{*}(-\overline{p}_{1}^{2})}{-\overline{p}_{1}^{2}}\frac{F^{*}(-\overline{p}_{2}^{2})}{-\overline{p}_{2}^{2}}\nonumber \\
 &  & \times\sum_{\textrm{spin of }l,\overline{l}}\left[u_{1\mu}u_{2\nu}L^{\mu\nu}(\overline{p}_{1},\overline{p}_{2};k_{1},k_{2})\right]\left[u_{1\sigma}u_{2\rho}L^{\sigma\rho*}(\overline{p}_{1},\overline{p}_{2};k_{1},k_{2})\right]\nonumber \\
 &  & \times(2\pi)^{4}\delta^{(4)}\left(\overline{p}_{1}+\overline{p}_{2}-k_{1}-k_{2}\right),\label{eq:cross-section-class-2}
\end{eqnarray}
where we have used $P_{1}^{\prime z}=P_{A1}^{z}$ as the solution
to the energy conservation equation with the conditions $\mathbf{P}_{1T}=\mathbf{P}_{1T}^{\prime}$
and $\mathbf{P}_{2T}=\mathbf{P}_{2T}^{\prime}$. The result of (\ref{eq:cross-section-class-2})
is consistent with the results in Ref. \citep{Vidovic:1992ik,Hencken:1994my}.

On the other hand, we can carry out the integrals over $p_{i}^{\prime}\cdot u_{i}$,
$p_{i}\cdot u_{i}$, $\omega_{i}^{\prime}$ and $\mathbf{P}_{iT}^{\prime}$
for $i=1,2$ in Eq. (\ref{eq:cross-section-2}), the result is 
\begin{eqnarray}
\sigma & = & \frac{Z^{4}e^{4}}{2\gamma^{4}v^{3}}\int d^{2}\mathbf{b}_{T}d^{2}\mathbf{b}_{1T}d^{2}\mathbf{b}_{2T}\int\frac{d\omega_{1}d^{2}\mathbf{p}_{1T}}{(2\pi)^{3}}\frac{d\omega_{2}d^{2}\mathbf{p}_{2T}}{(2\pi)^{3}}\nonumber \\
 &  & \times\int\frac{d^{2}\mathbf{P}_{1T}}{(2\pi)^{2}}\frac{d^{2}\mathbf{P}_{2T}}{(2\pi)^{2}}\phi_{T}(\mathbf{P}_{1T})\phi_{T}(\mathbf{P}_{2T})\delta^{(2)}\left(\mathbf{b}_{T}-\mathbf{b}_{1T}+\mathbf{b}_{2T}\right)\nonumber \\
 &  & \times\int\frac{d^{2}\mathbf{p}_{1T}^{\prime}}{(2\pi)^{2}}e^{-i\mathbf{b}_{1T}\cdot(\mathbf{p}_{1T}^{\prime}-\mathbf{p}_{1T})}\frac{F^{*}(-\overline{p}_{1}^{\prime2})}{-\overline{p}_{1}^{\prime2}}\frac{F(-\overline{p}_{1}^{2})}{-\overline{p}_{1}^{2}}\phi_{T}^{*}(\mathbf{P}_{1T}^{\prime})\nonumber \\
 &  & \times\int\frac{d^{2}\mathbf{p}_{2T}^{\prime}}{(2\pi)^{2}}e^{-i\mathbf{b}_{2T}\cdot(\mathbf{p}_{2T}^{\prime}-\mathbf{p}_{2T})}\frac{F^{*}(-\overline{p}_{2}^{\prime2})}{-\overline{p}_{2}^{\prime2}}\frac{F(-\overline{p}_{2}^{2})}{-\overline{p}_{2}^{2}}\phi_{T}^{*}(\mathbf{P}_{2T}^{\prime})G^{2}\left[(P_{1}^{\prime z}-P_{A1}^{z})^{2}\right]\nonumber \\
 &  & \times\int\frac{d^{3}k_{1}}{(2\pi)^{3}2E_{k1}}\frac{d^{3}k_{2}}{(2\pi)^{3}2E_{k2}}\sum_{\textrm{spin of }l,\overline{l}}\left[u_{1\mu}u_{2\nu}L^{\mu\nu}(\overline{p}_{1},\overline{p}_{2};k_{1},k_{2})\right]\nonumber \\
 &  & \times\left[u_{1\sigma}u_{2\rho}L^{\sigma\rho*}(\overline{p}_{1}^{\prime},\overline{p}_{2}^{\prime};k_{1},k_{2})\right](2\pi)^{4}\delta^{(4)}\left(\overline{p}_{1}+\overline{p}_{2}-k_{1}-k_{2}\right),\label{eq:cross-section-photon-impact}
\end{eqnarray}
where $\omega_{i}^{\prime}=\omega_{i}-E_{Pi}+E_{Pi^{\prime}}$, $\mathbf{P}_{1T}^{\prime}=\mathbf{P}_{1T}-\mathbf{p}_{1T}+\mathbf{p}_{1T}^{\prime}$
and $\mathbf{P}_{2T}^{\prime}=\mathbf{P}_{2T}-\mathbf{p}_{2T}+\mathbf{p}_{2T}^{\prime}$.
Note that $\mathbf{p}_{1T}^{\prime}$ and $\mathbf{p}_{2T}^{\prime}$
are now free variables. We see in the above formula that the couplings
of impact parameters with nuclear momenta have been converted to those
with photon momenta by integration over nuclear transverse momenta
$\mathbf{P}_{1T}^{\prime}$ and $\mathbf{P}_{2T}^{\prime}$. Completing
the integrals over $\mathbf{b}_{1T}$ and $\mathbf{b}_{2T}$ will
gives a term 
\begin{equation}
e^{-i\mathbf{b}_{T}\cdot(\mathbf{p}_{1T}^{\prime}-\mathbf{p}_{1T})}\delta^{(2)}\left(\mathbf{p}_{1T}+\mathbf{p}_{2T}+\mathbf{p}_{1T}^{\prime}-\mathbf{p}_{2T}^{\prime}\right),
\end{equation}
in the inetgrand.

If we make a similar ansatz to (\ref{eq:wave_ansatz}) for transverse
momentum amplitudes we can simplify Eq. (\ref{eq:cross-section-photon-impact})
significantly,  
\begin{eqnarray}
 &  & \phi_{T}(\mathbf{P}_{1T})\phi_{T}^{*}(\mathbf{P}_{1T}^{\prime})\phi_{T}(\mathbf{P}_{2T})\phi_{T}^{*}(\mathbf{P}_{2T}^{\prime})\nonumber \\
 & \approx & (2\pi)^{4}\delta^{(2)}\left[(\mathbf{P}_{1T}+\mathbf{P}_{1T}^{\prime})/2\right]\delta^{(2)}\left[(\mathbf{P}_{2T}+\mathbf{P}_{2T}^{\prime})/2\right]\nonumber \\
 &  & \times G_{T}\left[(\mathbf{P}_{1T}^{\prime}-\mathbf{P}_{1T})^{2}\right]G_{T}\left[(\mathbf{P}_{2T}^{\prime}-\mathbf{P}_{2T})^{2}\right]\nonumber \\
 & = & (2\pi)^{4}\delta^{(2)}\left[\mathbf{P}_{1T}-(\mathbf{p}_{1T}-\mathbf{p}_{1T}^{\prime})/2\right]\delta^{(2)}\left[\mathbf{P}_{2T}-(\mathbf{p}_{2T}-\mathbf{p}_{2T}^{\prime})/2\right]\nonumber \\
 &  & \times G_{T}\left[(\mathbf{p}_{1T}^{\prime}-\mathbf{p}_{1T})^{2}\right]G_{T}\left[(\mathbf{p}_{2T}^{\prime}-\mathbf{p}_{2T})^{2}\right],
\end{eqnarray}
where $G_{T}(x^{2})$ may differ from $G(x^{2})$ for longitudinal
momenta but with similar behavior: it is a positive function with
$G_{T}(0)=1$ and decreases rapidly with growing $x^{2}$. Then the
integrals over $\mathbf{P}_{1T}$ and $\mathbf{P}_{2T}$ in Eq. (\ref{eq:cross-section-photon-impact})
can be completed and we obtain
\begin{eqnarray}
\sigma & = & \frac{Z^{4}e^{4}}{2\gamma^{4}v^{3}}\int d^{2}\mathbf{b}_{T}d^{2}\mathbf{b}_{1T}d^{2}\mathbf{b}_{2T}\delta^{(2)}\left(\mathbf{b}_{T}-\mathbf{b}_{1T}+\mathbf{b}_{2T}\right)\nonumber \\
 &  & \times\int\frac{d\omega_{1}d^{2}\mathbf{p}_{1T}}{(2\pi)^{3}}\frac{d\omega_{2}d^{2}\mathbf{p}_{2T}}{(2\pi)^{3}}\nonumber \\
 &  & \times\int\frac{d^{2}\mathbf{p}_{1T}^{\prime}}{(2\pi)^{2}}e^{-i\mathbf{b}_{1T}\cdot(\mathbf{p}_{1T}^{\prime}-\mathbf{p}_{1T})}\frac{F^{*}(-\overline{p}_{1}^{\prime2})}{-\overline{p}_{1}^{\prime2}}\frac{F(-\overline{p}_{1}^{2})}{-\overline{p}_{1}^{2}}G_{T}\left[(\mathbf{p}_{1T}^{\prime}-\mathbf{p}_{1T})^{2}\right]\nonumber \\
 &  & \times\int\frac{d^{2}\mathbf{p}_{2T}^{\prime}}{(2\pi)^{2}}e^{-i\mathbf{b}_{2T}\cdot(\mathbf{p}_{2T}^{\prime}-\mathbf{p}_{2T})}\frac{F^{*}(-\overline{p}_{2}^{\prime2})}{-\overline{p}_{2}^{\prime2}}\frac{F(-\overline{p}_{2}^{2})}{-\overline{p}_{2}^{2}}G_{T}\left[(\mathbf{p}_{2T}^{\prime}-\mathbf{p}_{2T})^{2}\right]\nonumber \\
 &  & \times\int\frac{d^{3}k_{1}}{(2\pi)^{3}2E_{k1}}\frac{d^{3}k_{2}}{(2\pi)^{3}2E_{k2}}\sum_{\textrm{spin of }l,\overline{l}}\left[u_{1\mu}u_{2\nu}L^{\mu\nu}(\overline{p}_{1},\overline{p}_{2};k_{1},k_{2})\right]\nonumber \\
 &  & \times\left[u_{1\sigma}u_{2\rho}L^{\sigma\rho*}(\overline{p}_{1}^{\prime},\overline{p}_{2}^{\prime};k_{1},k_{2})\right](2\pi)^{4}\delta^{(4)}\left(\overline{p}_{1}+\overline{p}_{2}-k_{1}-k_{2}\right),\label{eq:cs-photon-impact}
\end{eqnarray}
where we have used $P_{1}^{\prime z}=P_{A1}^{z}$ as the solution
to the energy conservation equation with the conditions $\mathbf{P}_{1T}=-\mathbf{P}_{1T}^{\prime}$
and $\mathbf{P}_{2T}=-\mathbf{P}_{2T}^{\prime}$, and $\omega_{i}^{\prime}=\omega_{i}$
for $i=1,2$.

We emphasize that one of our main results in this work is the cross
section (\ref{eq:cs-photon-impact}) with impact parameter dependence
which encodes the information of photons in transverse phase space.
This is the basis for the derivation of EPA results in Sect. \ref{sec:Connection-to-EPA}
at the relativistic limit. We will show that Eq. (\ref{eq:cs-photon-impact})
contains all high order contributions of $\gamma\gamma\rightarrow l\overline{l}$
at the tree level. In Sect. \ref{sec:Numerical-results}, we will
compute cross sections based on Eq. (\ref{eq:cs-photon-impact}) and
compare our results with experimental data.

We will also parametrize the TMD correlation function $S_{\mu\nu}$
in Eq. (\ref{eq:TMD_unpol_01}) and implement the TMD factorization
formalism in Sect. \ref{sec:TMD-factorization-formalism}. 


\section{Connection to EPA \label{sec:Connection-to-EPA}}

In this section, we will derive the EPA result from Eq. (\ref{eq:cs-photon-impact}).
Now we evaluate $u_{1\mu}u_{2\nu}L^{\mu\nu}(\overline{p}_{1},\overline{p}_{2};k_{1},k_{2})$
and $u_{1\sigma}u_{2\rho}L^{\sigma\rho*}(\overline{p}_{1}^{\prime},\overline{p}_{2}^{\prime};k_{1},k_{2})$
in Eq. (\ref{eq:cross-section-photon-impact}). In order to simply
notations, from now on, we resume the use of $p_{1}$, $p_{2}$, $p_{1}^{\prime}$
and $p_{2}^{\prime}$ in $L^{\mu\nu}$ and $L^{\sigma\rho*}$ for
$\overline{p}_{1}$, $\overline{p}_{2}$, $\overline{p}_{1}^{\prime}$
and $\overline{p}_{2}^{\prime}$ respectively if there is no ambiguity.
It is convenient to rewrite $u_{1\mu}u_{2\nu}\Gamma^{\mu\nu}(p_{1},p_{2};k_{1}k_{2})$
in light-cone coordinates 
\begin{eqnarray}
u_{1\mu}u_{2\nu}L^{\mu\nu}(p_{1},p_{2};k_{1}k_{2}) & = & \gamma^{2}v^{2}\frac{p_{1}^{i}}{\omega_{1}}\frac{p_{2}^{j}}{\omega_{2}}L^{ij}\nonumber \\
 &  & -2\gamma^{2}v^{2}\left(\frac{p_{1}^{i}}{\omega_{1}}\frac{p_{2}^{+}}{\omega_{2}}L^{i-}+\frac{p_{1}^{-}}{\omega_{1}}\frac{p_{2}^{j}}{\omega_{2}}L^{+j}\right)\nonumber \\
 &  & +4\gamma^{2}v^{2}\frac{p_{1}^{-}}{\omega_{1}}\frac{p_{2}^{+}}{\omega_{2}}L^{+-},\label{eq:approx_01}
\end{eqnarray}
where $i,j=x,y$ stand for transverse directions, we have used Eq.
(\ref{eq:relation_Gamma_01}), and $p_{1,2}^{\pm}$ are given by

\begin{equation}
p_{1}^{\pm}=\frac{\omega_{1}}{\sqrt{2}}\left(1\pm\frac{1}{v}\right),\;\;p_{2}^{\pm}=\frac{\omega_{2}}{\sqrt{2}}\left(1\mp\frac{1}{v}\right).
\end{equation}
At the relativistic limit $v\rightarrow1$ and $\gamma\rightarrow\infty$,
the following power counting rules hold in light-cone coordinate 
\begin{equation}
\frac{p_{1}^{+}}{\omega_{1}},\frac{p_{2}^{-}}{\omega_{2}}\sim\mathcal{O}(1),\qquad\frac{p_{1}^{-}}{\omega_{1}},\frac{p_{2}^{+}}{\omega_{2}}\sim\mathcal{O}(\gamma^{-2}),\qquad\frac{\mathbf{p}_{1}^{T}}{\omega_{1}},\frac{\mathbf{p}_{2}^{T}}{\omega_{2}}\sim\mathcal{O}(\gamma^{-1}).\label{eq:power-counting}
\end{equation}
It implies that photons are almost on-shell \citep{Vidovic:1992ik},
\begin{equation}
\frac{p^{2}}{\omega^{2}}\sim\mathcal{O}(\gamma^{-2}),\label{eq:power_01}
\end{equation}
Following (\ref{eq:power-counting}), the first, second and third
term of (\ref{eq:approx_01}) are $O(1)$, $O(\gamma^{-1})$ and $O(\gamma^{-2})$
respectively. Therefore the leading order contribution comes from
the first term. The expansion in (\ref{eq:approx_01}) can also be
interpreted as the photon virtuality expansion. Details can be found
in Appendix \ref{sec:Lightcone}.

At the relativistic limit, the cross section (\ref{eq:cs-photon-impact})
can be put into a compact form 
\begin{equation}
\sigma=\sigma_{0}+\delta\sigma,\label{eq:total_cross_EPA}
\end{equation}
where $\sigma_{0}$ is the leading order contribution from the first
term of Eq. (\ref{eq:approx_01}) 
\begin{eqnarray}
\gamma^{2}v^{2}\frac{p_{1}^{i}p_{2}^{j}}{\omega_{1}\omega_{2}}L^{ij} & \approx & \gamma^{2}v^{2}\frac{|\mathbf{p}_{1T}||\mathbf{p}_{2T}|}{\omega_{1}\omega_{2}}\widehat{p}_{1}^{i}\widehat{p}_{2}^{j}L^{ij}(p_{1},p_{2};k_{1},k_{2}),\nonumber \\
\gamma^{2}v^{2}\frac{p_{1}^{\prime k}p_{2}^{\prime l}}{\omega_{1}^{\prime}\omega_{2}^{\prime}}L^{kl} & \approx & \gamma^{2}v^{2}\frac{|\mathbf{p}_{1T}^{\prime}||\mathbf{p}_{2T}^{\prime}|}{\omega_{1}^{\prime}\omega_{2}^{\prime}}\widehat{p}_{1}^{\prime k}\widehat{p}_{2}^{\prime l}L^{kl}(p_{1}^{\prime},p_{2}^{\prime};k_{1},k_{2}),\label{eq:leading-term}
\end{eqnarray}
and $\delta\sigma$ represents the corrections from other terms. In
the above formula, $i,j,k,l$ are indices of two transverse directions,
and $\widehat{p}_{1,2}^{i}$ and $\widehat{p}_{1,2}^{\prime i}$ denote
the directions (unit vectors) of $\mathbf{p}_{1T,2T}$ and $\mathbf{p}_{1T,2T}^{\prime}$
respectively which play the role of polarization vectors of photons
so the summation over $i,j,k,l$ represents that over photon polarizations.
Therefore we can define the last two lines of Eq. (\ref{eq:cs-photon-impact})
as a kind of cross section of the photon fusion to produce dileptons
\begin{eqnarray}
\sigma_{\gamma\gamma\rightarrow l\overline{l}}(p_{1},p_{2};p_{1}^{\prime},p_{2}^{\prime}) & = & \frac{1}{8\omega_{1}^{\prime}\omega_{2}^{\prime}}\int\frac{d^{3}k_{1}}{(2\pi)^{3}2E_{k1}}\frac{d^{3}k_{2}}{(2\pi)^{3}2E_{k2}}\sum_{\textrm{spin of }l,\overline{l}}\left[\widehat{p}_{1}^{i}\widehat{p}_{2}^{j}L^{ij}(p_{1},p_{2};k_{1},k_{2})\right]\nonumber \\
 &  & \times\left[\widehat{p}_{1}^{\prime k}\widehat{p}_{2}^{\prime l}L^{kl}(p_{1}^{\prime},p_{2}^{\prime};k_{1},k_{2})\right](2\pi)^{4}\delta^{(4)}\left(p_{1}+p_{2}-k_{1}-k_{2}\right).
\end{eqnarray}
When $p_{1}=p_{1}^{\prime}$ and $p_{2}=p_{2}^{\prime}$, $\sigma_{\gamma\gamma\rightarrow l\overline{l}}(p_{1}p_{2},p_{1}^{\prime}p_{2}^{\prime})$
becomes the ordinary cross section for production of lepton pairs
by two photons

The leading order cross secction from Eq. (\ref{eq:cs-photon-impact})
is put into the form

\begin{eqnarray}
\sigma_{0}(A_{1}A_{2}\rightarrow l\overline{l}) & \approx & \int d^{2}\mathbf{b}_{T}d^{2}\mathbf{b}_{1T}d^{2}\mathbf{b}_{2T}\delta^{(2)}\left(\mathbf{b}_{T}-\mathbf{b}_{1T}+\mathbf{b}_{2T}\right)\nonumber \\
 &  & \times\int d\omega_{1}d^{2}\mathbf{p}_{1T}d\omega_{2}d^{2}\mathbf{p}_{2T}\nonumber \\
 &  & \times\frac{Z^{2}\alpha}{\omega_{1}\pi^{2}}\int\frac{d^{2}\mathbf{p}_{1T}^{\prime}}{(2\pi)^{2}}e^{-i\mathbf{b}_{1T}\cdot(\mathbf{p}_{1T}^{\prime}-\mathbf{p}_{1T})}|\mathbf{p}_{1T}||\mathbf{p}_{1T}^{\prime}|\nonumber \\
 &  & \times\frac{F^{*}(-p_{1}^{\prime2})}{-p_{1}^{\prime2}}\frac{F(-p_{1}^{2})}{-p_{1}^{2}}G_{T}\left[(\mathbf{p}_{1T}^{\prime}-\mathbf{p}_{1T})^{2}\right]\nonumber \\
 &  & \times\frac{Z^{2}\alpha}{\omega_{1}\pi^{2}}\int\frac{d^{2}\mathbf{p}_{2T}^{\prime}}{(2\pi)^{2}}e^{-i\mathbf{b}_{2T}\cdot(\mathbf{p}_{2T}^{\prime}-\mathbf{p}_{2T})}|\mathbf{p}_{2T}||\mathbf{p}_{2T}^{\prime}|\nonumber \\
 &  & \times\frac{F^{*}(-p_{2}^{\prime2})}{-p_{2}^{\prime2}}\frac{F(-p_{2}^{2})}{-p_{2}^{2}}G_{T}\left[(\mathbf{p}_{2T}^{\prime}-\mathbf{p}_{2T})^{2}\right]\nonumber \\
 &  & \times\sigma_{\gamma\gamma\rightarrow l\overline{l}}(p_{1},p_{2};p_{1}^{\prime},p_{2}^{\prime}).\label{eq:total_cross_04}
\end{eqnarray}
If we approximate $\mathbf{p}_{iT}^{\prime}\approx\mathbf{p}_{iT}$
for $i=1,2$, we have 
\begin{eqnarray}
G_{T}\left[(\mathbf{p}_{iT}^{\prime}-\mathbf{p}_{iT})^{2}\right] & \approx & 1,\nonumber \\
\sigma_{\gamma\gamma\rightarrow l\overline{l}}(p_{1},p_{2};p_{1}^{\prime},p_{2}^{\prime}) & \approx & \sigma_{\gamma\gamma\rightarrow l\overline{l}}(p_{1},p_{2})\nonumber \\
 & \approx & \sigma_{\gamma\gamma\rightarrow l\overline{l}}(\omega_{1},\omega_{2}).
\end{eqnarray}
Then Eq. (\ref{eq:total_cross_04}) can be written in a compact way
\begin{eqnarray}
\sigma_{0}(A_{1}A_{2}\rightarrow l\overline{l}) & \approx & \int d^{2}\mathbf{b}_{T}d^{2}\mathbf{b}_{1T}d^{2}\mathbf{b}_{2T}\int d\omega_{1}d^{2}\mathbf{p}_{1T}d\omega_{2}d^{2}\mathbf{p}_{2T}\nonumber \\
 &  & \times n_{A1}\left(\omega_{1},\mathbf{b}_{1T},\mathbf{p}_{1T}\right)n_{A2}\left(\omega_{2},\mathbf{b}_{2T},\mathbf{p}_{2T}\right)\nonumber \\
 &  & \times\delta^{(2)}\left(\mathbf{b}_{T}-\mathbf{b}_{1T}+\mathbf{b}_{2T}\right)\sigma_{\gamma\gamma\rightarrow l\overline{l}}(\omega_{1},\omega_{2})\nonumber \\
 & = & \int d^{2}\mathbf{b}_{T}d^{2}\mathbf{b}_{1T}d^{2}\mathbf{b}_{2T}\int d\omega_{1}d\omega_{2}\nonumber \\
 &  & \times n_{A1}\left(\omega_{1},\mathbf{b}_{1T}\right)n_{A2}\left(\omega_{2},\mathbf{b}_{2T}\right)\nonumber \\
 &  & \times\delta^{(2)}\left(\mathbf{b}_{T}-\mathbf{b}_{1T}+\mathbf{b}_{2T}\right)\sigma_{\gamma\gamma\rightarrow l\overline{l}}(\omega_{1},\omega_{2}),\label{eq:sigma-0}
\end{eqnarray}
where the photon flux spectra from $A_{1}$ and $A_{2}$ are

\begin{eqnarray}
n_{Ai}\left(\omega_{i},\mathbf{b}_{iT},\mathbf{p}_{iT}\right) & \approx & \frac{Z^{2}\alpha}{\omega_{i}\pi^{2}}\int\frac{d^{2}\mathbf{p}_{iT}^{\prime}}{(2\pi)^{2}}|\mathbf{p}_{iT}||\mathbf{p}_{iT}^{\prime}|e^{-i\mathbf{b}_{iT}\cdot(\mathbf{p}_{iT}^{\prime}-\mathbf{p}_{iT})}\nonumber \\
 &  & \times\frac{F^{*}(-p_{i}^{\prime2})}{-p_{i}^{\prime2}}\frac{F(-p_{i}^{2})}{-p_{i}^{2}},\nonumber \\
n_{Ai}\left(\omega_{i},\mathbf{b}_{iT}\right) & = & \int d^{2}\mathbf{p}_{iT}n_{Ai}\left(\omega_{i},\mathbf{b}_{iT},\mathbf{p}_{iT}\right)\nonumber \\
 & = & \frac{4Z^{2}\alpha}{\omega_{i}}\left|\int\frac{d^{2}\mathbf{p}_{iT}}{\left(2\pi\right)^{2}}e^{i\mathbf{b}_{iT}\cdot\mathbf{p}_{iT}}\mathbf{p}_{iT}\frac{F(-p_{i}^{2})}{-p_{i}^{2}}\right|^{2},
\end{eqnarray}
for $i=1,2$. The impact parameter structure in Eq. (\ref{eq:sigma-0})
is similar to the space-dependent photon flux defined in Ref. \citep{Xiao:2020ddm,Klein:2020jom,Shi:2020djm}.
The approximation that leads to the result of (\ref{eq:sigma-0})
is called the generalized EPA or gEPA \citep{Vidovic:1992ik}.

We can complete the integrals over $\mathbf{b}_{T}$, $\mathbf{b}_{1T}$,
and $\mathbf{b}_{2T}$ in Eq. (\ref{eq:sigma-0}) to obtain 
\begin{equation}
\sigma_{0}(A_{1}A_{2}\rightarrow l\overline{l})=\int d\omega_{1}d\omega_{2}n_{A1}(\omega_{1})n_{A2}(\omega_{2})\sigma_{\gamma\gamma\rightarrow l\overline{l}}(\omega_{1},\omega_{2}),\label{eq:leading-o-cr}
\end{equation}
where $n_{A1}$ and $n_{A2}$ are the photon fluxes generated by $A_{1}$
and $A_{2}$ respectively and are given by 
\begin{equation}
n_{Ai}(\omega_{i})=\frac{4Z^{2}\alpha}{\omega_{i}}\int\frac{d^{2}\mathbf{p}_{iT}}{(2\pi)^{2}}\mathbf{p}_{iT}^{2}\left|\frac{F(-p_{i}^{2})}{-p_{i}^{2}}\right|^{2},\label{eq:photon-flux}
\end{equation}
for $i=1,2$. Note that the photons are very close to real ones as
shown in Eq. (\ref{eq:power_01}). The approximation that leads to
the result of (\ref{eq:leading-o-cr}) is called EPA.

About $\delta\sigma$ in (\ref{eq:total_cross_EPA}), the cross section
beyond the EPA, with the same approximation as for Eq. (\ref{eq:sigma-0}),
we obtain 
\begin{eqnarray}
\delta\sigma & = & \frac{Z^{4}\alpha^{2}v}{8\pi^{4}}\int\frac{d^{3}k_{1}}{(2\pi)^{3}2E_{k_{1}}}\frac{d^{3}k_{2}}{(2\pi)^{3}2E_{k_{2}}}\int\frac{d\omega_{1}}{\omega_{1}^{2}}\frac{d\omega_{2}}{\omega_{2}^{2}}\int d^{2}\mathbf{p}_{1T}d^{2}\mathbf{p}_{2T}\nonumber \\
 &  & \times\left|\frac{F(-p_{1}^{2})}{-p_{1}^{2}}\right|^{2}\left|\frac{F(-p_{2}^{2})}{-p_{2}^{2}}\right|^{2}(2\pi)^{4}\delta^{4}(p_{1}+p_{2}-k_{1}-k_{2})\mathcal{I},\label{eq:total_cross_02-1}
\end{eqnarray}
where $\mathcal{I}$ is defined through 
\begin{equation}
u_{1\mu}u_{2\nu}L^{\mu\nu}u_{1\sigma}u_{2\rho}L^{\sigma\rho*}=\frac{\gamma^{4}v^{4}}{\omega_{1}^{2}\omega_{2}^{2}}\left[p_{1}^{i}p_{2}^{j}p_{1}^{k}p_{2}^{l}L^{ij}L^{\sigma\rho*}+\mathcal{I}\right],
\end{equation}
where $L^{\mu\nu}\equiv L^{\mu\nu}(p_{1},p_{2};k_{1},k_{2})$ and
$L^{\sigma\rho*}\equiv(p_{1},p_{2};k_{1},k_{2})$ which are given
by (\ref{eq:l-mu-nu-2}). Note that the first term inside the square
brackets gives $\sigma_{0}$. The explicit form of $\mathcal{I}$
is given by 
\begin{eqnarray}
\mathcal{I} & = & -2p_{1}^{i}p_{2}^{j}L^{ij}\left(p_{2}^{+}p_{1}^{k}L^{k-*}+p_{1}^{-}p_{2}^{k}L^{+k*}\right)\nonumber \\
 &  & +4p_{1}^{i}p_{2}^{j}p_{1}^{-}p_{2}^{+}L^{ij}L^{+-*}\nonumber \\
 &  & -2\left(p_{2}^{+}p_{1}^{i}L^{i-}+p_{1}^{-}p_{2}^{i}L^{+i}\right)p_{1}^{k}p_{2}^{l}L^{kl*}\nonumber \\
 &  & +4\left(p_{2}^{+}p_{1}^{i}L^{i-}+p_{1}^{-}p_{2}^{j}L^{+j}\right)\left(p_{2}^{+}p_{1}^{k}L^{k-*}+p_{1}^{-}p_{2}^{k}L^{+k*}\right)\nonumber \\
 &  & -8\left(p_{2}^{+}p_{1}^{i}L^{i-}+p_{1}^{-}p_{2}^{j}L^{+j}\right)p_{1}^{-}p_{2}^{+}L^{+-*}\nonumber \\
 &  & +4p_{1}^{-}p_{2}^{+}p_{1}^{k}p_{2}^{l}L^{kl*}L^{+-}\nonumber \\
 &  & -8p_{1}^{-}p_{2}^{+}L^{+-}\left(p_{2}^{+}p_{1}^{k}L^{k-*}+p_{1}^{-}p_{2}^{k}L^{+k*}\right)\nonumber \\
 &  & +16(p_{1}^{-})^{2}(p_{2}^{+})^{2}L^{+-}L^{+-*}.\label{eq:expression_I_01}
\end{eqnarray}
So we see that the cross section (\ref{eq:cross-section-photon-impact})
contains not only the result of EPA, but also the result beyond the
EPA.


\section{Connection to TMD factorization formalism \label{sec:TMD-factorization-formalism}}

In this section, we will discuss the connection of our results to
those from the TMD factorization formalism. Here we just give a concise
comparison between the results from our approach and those from the
TMD factorization formalism. The details of the TMD factorization
formalism applied to UPC can be found in Refs. \citep{Klein:2018fmp,Li:2019yzy,Li:2019sin,Ringer:2019rfk}.
The recent developments in the impact-parameter dependent cross sections
in terms of photon Wigner functions are given in Ref. \citep{Xiao:2020ddm,Klein:2020jom}.

We work in the light-front formalism and discuss photon Wigner functions
based on Ref. \citep{Klein:2020jom,Xiao:2020ddm}. The gauge invariant
TMD correlation function of EM fields \citep{Mulders:2000sh} for
an unpolarized nucleus moving in $+z$ direction is defined as 
\begin{equation}
\mathcal{W}^{\mu\nu;\rho\sigma}(P_{1},p_{1})=\int\frac{d^{4}\xi}{(2\pi)^{4}}e^{ip_{1}\cdot\xi}\left\langle P_{1}\right|F^{\mu\nu}(0)F^{\rho\sigma}(\xi)\left|P_{1}\right\rangle .
\end{equation}
Similarly one can define the TMD correlation function $\mathcal{W}^{\mu\nu;\rho\sigma}(P_{2},p_{2})$
for the nucleus moving in $-z$ direction. In Sect. \ref{sec:Classical-field-approximation}
we use the Lorentz gauge. Both $\mathcal{W}^{\mu\nu;\rho\sigma}$
and the cross section are gauge invariant, therefore, in this section
we follow the standard TMD factorization formalism to choose the light-cone
gauge, $A^{\pm}=0$, for nuclei 1 and 2 moving in $\pm z$ direction.
In the light-cone gauge the classical solutions to EM fields read
\begin{eqnarray}
A_{(1,2)}^{\mu}(p) & = & 2\pi Z_{1,2}e\delta(p\cdot u_{1,2})\frac{F(-p^{2})}{-p^{2}}\left[u_{1,2}^{\mu}-\frac{p^{\mu}}{p^{\pm}}u_{1,2}^{\pm}\right],\label{eq:light-cone-a}
\end{eqnarray}
where the upper/lower sign corresponds to nuclei 1 and 2 moving in
$\pm z$ direction, as an comparison, the result in the Lorentz gauge
is given in Eq. (\ref{eq:classical_01}). One can verify that the
TMD correlation functions are related to the correlation functions
in Eq. (\ref{eq:TMD_unpol_01}) as 
\begin{eqnarray}
\mathcal{W}^{+\nu;+\sigma}(P_{1},p_{1}) & = & (p_{1}^{+})^{2}S^{\nu\sigma}(P_{1},p_{1}),\nonumber \\
\mathcal{W}^{+\nu;+\sigma}(P_{2},p_{2}) & = & (p_{2}^{-})^{2}S^{\nu\sigma}(P_{2},p_{2}).
\end{eqnarray}
Note that there is a different sign from Ref. \citep{Li:2019sin}.

We follow the twist expansion in the TMD factorization formalism \citep{Brodsky:1997de,Ji:2020ect,Collins:1981uw,Collins:1984kg,Tangerman:1994eh,Bacchetta:2008xw,Ji:2004wu,Mulders:2000sh},
where the twist-2 contribution comes from $\mathcal{W}^{+i;+j}$,
the twist-3 contribution comes from $\mathcal{W}^{+i;+-}$ and $\mathcal{W}^{ij;l+}$,
and $\mathcal{W}^{+-;+-}$ contributes at twist-4.

In the order of twist-2, we can parametrize correlation functions
for nuclei 1 and 2 of the same species as 
\begin{eqnarray}
 &  & \int dp_{1}^{-}\mathcal{W}^{+i;+j}(P_{1},p_{1})\nonumber \\
 & = & \frac{1}{2}x_{1}P_{1}^{+}\left[-g_{T}^{ij}f^{\gamma}(x_{1},\mathbf{p}_{1T}^{2})+\left(\frac{p_{1T}^{i}p_{1T}^{j}}{M^{2}}+\frac{\mathbf{p}_{1T}^{2}}{2M^{2}}g_{T}^{ij}\right)h_{T}^{\gamma}(x_{1},\mathbf{p}_{1T}^{2})\right],\nonumber \\
 &  & \int dp_{2}^{+}\mathcal{W}^{+i;+j}(P_{2},p_{2})\nonumber \\
 & = & \frac{1}{2}x_{2}P_{2}^{-}\left[-g_{T}^{ij}f^{\gamma}(x_{2},\mathbf{p}_{2T}^{2})+\left(\frac{p_{2T}^{i}p_{2T}^{j}}{M^{2}}+\frac{\mathbf{p}_{2T}^{2}}{2M^{2}}g_{T}^{ij}\right)h_{T}^{\gamma}(x_{2},\mathbf{p}_{2T}^{2})\right],
\end{eqnarray}
where $x_{1}=p_{1}^{+}/P_{1}^{+}$ and $x_{2}=p_{2}^{-}/P_{2}^{-}$
denote light-cone momentum fractions. The TMD distribution $f^{\gamma}(x_{1},\mathbf{p}_{1T}^{2})$
represents the usual unpolarized photon distribution, while $h_{T}^{\gamma}(x_{1},\mathbf{p}_{1T}^{2})$
represents the distribution function of linearly-polarized photons
in an unpolarized nucleus \citep{Mulders:2000sh}. The cross section
(\ref{eq:diff_05}) up to twist-2 can be expressed by 
\begin{eqnarray}
\sigma_{\mathrm{twist\:2}} & \approx & \frac{1}{32(2\pi)^{6}}\int d^{3}k_{1}d^{3}k_{2}\int dp_{1}^{+}d^{2}\mathbf{p}_{1T}dp_{2}^{-}d^{2}\mathbf{p}_{2T}\frac{1}{vE_{A1}E_{A2}E_{k1}E_{k2}}\nonumber \\
 &  & \times\frac{1}{2x_{1}P_{1}^{+}}\frac{p_{1T}^{\sigma}p_{1T}^{\mu}}{M^{2}}h_{T}^{\gamma}(x_{1},\mathbf{p}_{1T}^{2})\frac{1}{2x_{2}P_{2}^{-}}\frac{p_{2T}^{\rho}p_{2T}^{\nu}}{M^{2}}h_{T}^{\gamma}(x_{2},\mathbf{p}_{2T}^{2})\nonumber \\
 &  & \times L_{\mu\nu;\sigma\rho}\left(p_{1},p_{2};p_{1},p_{2};k_{1},k_{2}\right)(2\pi)^{4}\delta^{4}(p_{1}+p_{2}-k_{1}-k_{2})\label{eq:diff_05-2}
\end{eqnarray}
where we have assumed $E_{P1}\approx E_{A1}$ and $E_{P2}\approx E_{A2}$
and performed the integrals over $\mathbf{P}_{1T}$ and $\mathbf{P}_{2T}$.

The TMD distributions $f^{\gamma}(x,\mathbf{p}_{T}^{2})$ and $h_{T}^{\gamma}(x,\mathbf{p}_{T}^{2})$
are non-perturbative in nature and cannot be derived directly from
the perturbation theory. In the classical field approximation based
on the classical solution (\ref{eq:light-cone-a}) we obtain 
\begin{eqnarray}
xf^{\gamma}(x,\mathbf{p}_{T}^{2}) & = & \frac{Z^{2}\alpha}{\pi^{2}}\mathbf{p}_{T}^{2}\left|\frac{F(-p^{2})}{-p^{2}}\right|^{2},\nonumber \\
xh_{T}^{\gamma}(x,\mathbf{p}_{T}^{2}) & = & \frac{2Z^{2}\alpha}{\pi^{2}}M^{2}\left|\frac{F(-p^{2})}{-p^{2}}\right|^{2},\label{eq:f_h_01}
\end{eqnarray}
which are consistent with the photon flux $n_{A}(\omega,\mathbf{p}_{T})$
in Eqs. (\ref{eq:photon-flux}). The details for the derivation of
the above result can be found in Appendix \ref{sec:Correlation-functions-in}.
Inserting Eq. (\ref{eq:f_h_01}) into Eq. (\ref{eq:diff_05-2}) yields,
\begin{eqnarray}
\frac{d\sigma_{\mathrm{twist\:2}}}{d^{3}k_{1}d^{3}k_{2}} & = & \frac{1}{2(2\pi)^{10}}Z^{4}\alpha^{2}v\int d\omega_{1}d^{2}\mathbf{p}_{1T}d\omega_{2}d^{2}\mathbf{p}_{2T}\nonumber \\
 &  & \times\frac{1}{E_{k1}E_{k2}}\frac{p_{1T}^{\sigma}p_{1T}^{\mu}p_{2T}^{\sigma}p_{2T}^{\mu}}{\omega_{1}^{2}\omega_{2}^{2}}\left|\frac{F(-p_{1}^{2})}{-p_{1}^{2}}\right|^{2}\left|\frac{F(-p_{2}^{2})}{-p_{2}^{2}}\right|^{2}\nonumber \\
 &  & \times L_{\mu\nu;\sigma\rho}\left(p_{1},p_{2};p_{1},p_{2};k_{1},k_{2}\right)(2\pi)^{4}\delta^{4}(p_{1}+p_{2}-k_{1}-k_{2})\label{eq:diff_05-3}
\end{eqnarray}
It is straightforward to prove that the above result is equivalent
to Eq. (\ref{eq:sigma-0}).

In the classical field approximation, the twist-3 correlation functions
are given by 
\begin{eqnarray}
\int dp_{1}^{-}\mathcal{W}^{+i;+-}(P_{1},p_{1}) & = & 2P_{1}^{+}p_{1}^{-}p_{1}^{i}\frac{Z^{2}\alpha}{\pi^{2}}\left|\frac{F(-p_{1}^{2})}{-p_{1}^{2}}\right|^{2},\nonumber \\
\int dp_{1}^{-}\mathcal{W}^{ij;k+}(P_{1},p_{1}) & = & 0,
\end{eqnarray}
and the twist-4 correlation function is 
\begin{equation}
\int dp_{1}^{-}\mathcal{W}^{+-;+-}(P_{1},p_{1})=4P_{1}^{+}(p_{1}^{-})^{2}\frac{Z^{2}\alpha}{\pi^{2}}\left|\frac{F(-p_{1}^{2})}{-p_{1}^{2}}\right|^{2}.
\end{equation}
No higher order (twist-3 and twist-4) correlation functions contribute
to the tree level diagram of $\gamma\gamma\rightarrow l\overline{l}$.

The total cross section in the TMD factorization formalism reads,
\begin{equation}
\sigma=\sigma_{\mathrm{twist\:2}}+\sigma_{\mathrm{twist}\:n},\label{eq:cross_TMD_01}
\end{equation}
where 
\begin{eqnarray}
\sigma_{\mathrm{twist}\:n} & = & \frac{Z^{4}\alpha^{2}v}{8\pi^{4}}\int\frac{d^{3}k_{1}}{(2\pi)^{3}2E_{k1}}\frac{d^{3}k_{2}}{(2\pi)^{3}2E_{k2}}\frac{d\omega_{1}}{\omega_{1}^{2}}\frac{d\omega_{2}}{\omega_{2}^{2}}d^{2}p_{1T}d^{2}p_{2T}\nonumber \\
 &  & \times\left|\frac{F(-p_{1}^{2})}{-p_{1}^{2}}\right|^{2}\left|\frac{F(-p_{2}^{2})}{-p_{2}^{2}}\right|^{2}(2\pi)^{4}\delta^{4}(p_{1}+p_{2}-k_{1}-k_{2})\mathcal{I},
\end{eqnarray}
with $\mathcal{I}$ given by Eq. (\ref{eq:expression_I_01}). It is
remarkably that $\sigma_{\mathrm{twist}\:n}$ is exactly $\delta\sigma$
in Eq. (\ref{eq:total_cross_02-1}). Therefore, since we have already
shown that $\sigma_{\mathrm{twist\:2}}=\sigma_{0}$ in Eq. (\ref{eq:diff_05-3}),
the total cross section (\ref{eq:cross_TMD_01}) in the TMD factorization
formalism is consistent with the EPA result (\ref{eq:total_cross_EPA}).

Thus we have shown the equivalence of EPA and TMD factorization formalism.
We would like to emphasis that our main result (\ref{eq:cs-photon-impact})
includes corrections beyond the EPA or twist-2. Note that, in principle,
one also needs to consider the Sudakov factor in the TMD factorization
theorem \citep{Klein:2018fmp,Li:2019sin,Li:2019yzy,Xiao:2020ddm}.
It is still a chanllenge to add the effective Sudakov factor into
our framework, which we reserve for a future study.


\begin{table}
\caption{The total cross sections from STAR measurements and theoretical models.
The numerical integration errors are labeled as ``int.''. \label{tab:Total_cross_section_01-1}}

\centering{}%
\begin{tabular}{c|c}
\hline 
Data or models & Total cross sections\tabularnewline
\hline 
\hline 
STAR data \citep{Adam:2019mby} & 0.261$\pm$0.004 (stat.)$\pm$0.013 (sys.)$\pm$0.034 (scale) mb\tabularnewline
\hline 
STARLight \citep{Klein:2016yzr} & $0.22$ mb\tabularnewline
\hline 
Zha et al's gEPA \citep{Zha:2018tlq} & $0.26$ mb\tabularnewline
\hline 
Zha et al's work \citep{Zha:2018tlq} & $0.26$ mb\tabularnewline
\hline 
This work Eq. (\ref{eq:cs-photon-impact}) & 0.252$\pm$0.0016 (int.) mb\tabularnewline
\hline 
gEPA Eq. (\ref{eq:sigma-0}) & 0.256$\pm$0.0030 (int.) mb\tabularnewline
\hline 
\end{tabular}
\end{table}

\begin{figure}
\centering{}\includegraphics[scale=0.4]{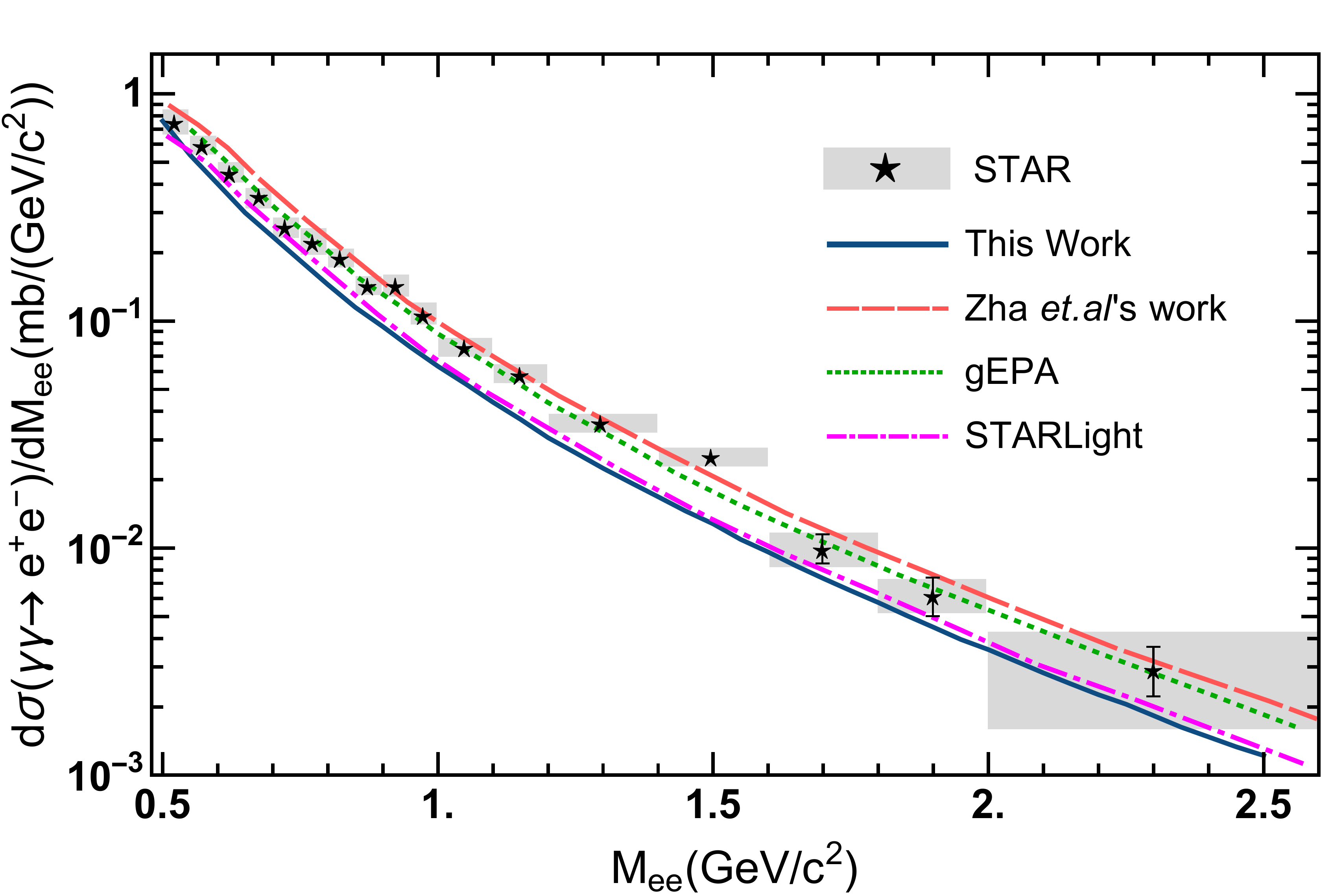}\caption{Differential cross sections as functions of the invariant mass of
the lepton pair $M_{ee}$. The blue-solid line represents our results
using Eq. (\ref{eq:cs-photon-impact}). The red-dashed, green-dotted
and magenta-dot-dashed lines represent the results from Ref. \citep{Zha:2018ywo}
based on QED, gEPA \citep{Zha:2018ywo} and STARLight \citep{Klein:2016yzr},
respectively. The points are STAR data \citep{Adam:2019mby}, while
the shaded areas stand for the experimental uncertainty. \label{fig:Mee}}
\end{figure}

\begin{figure}
\centering{}\includegraphics[scale=0.4]{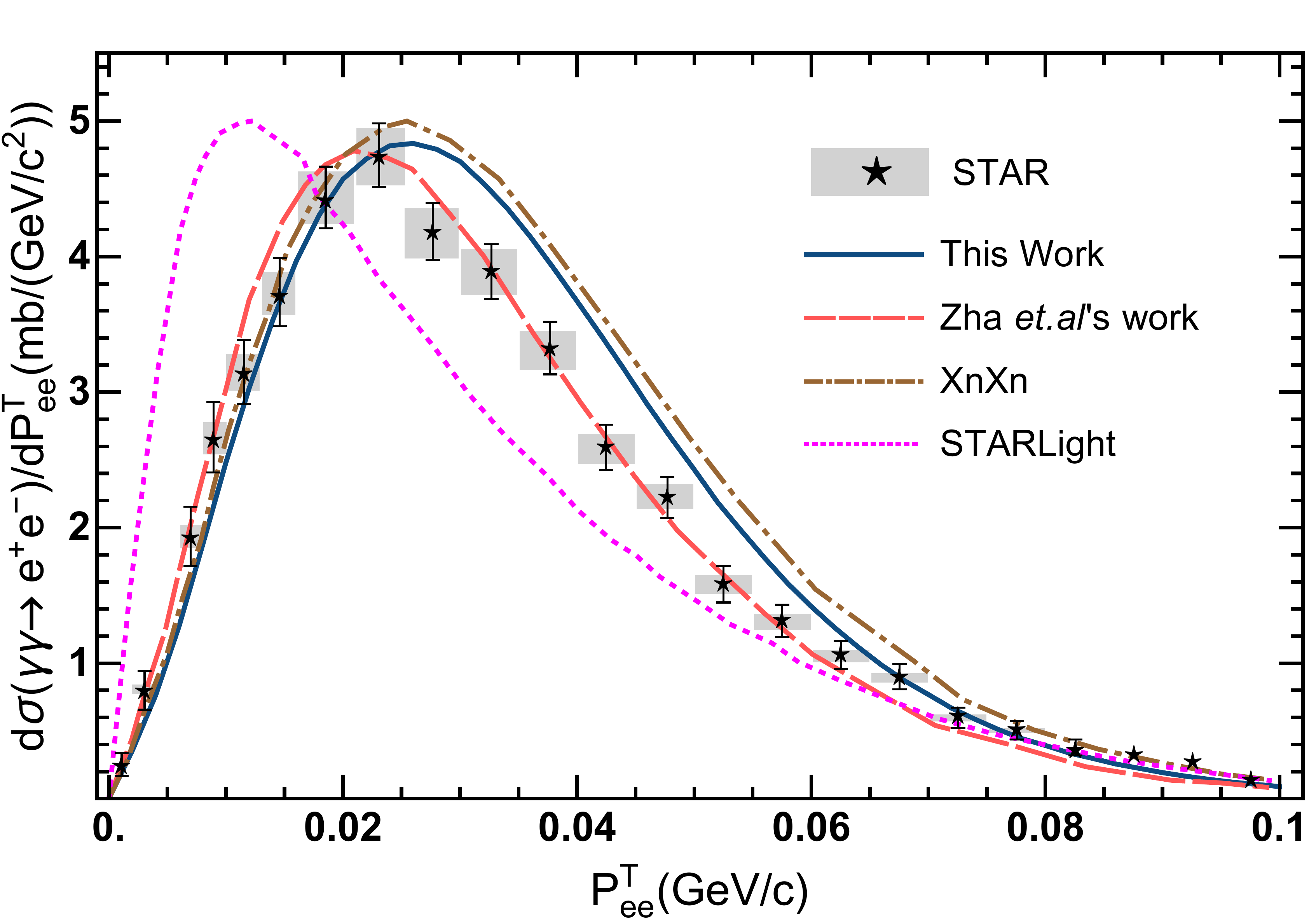}\caption{Differential cross sections as functions of lepton pair's transverse
momenta. The blue-solid line are our results using Eq. (\ref{eq:cs-photon-impact}).
The red-dashed, brown-dot-dashed and magenta-dotted lines are the
results from Ref. \citep{Zha:2018ywo} (times a factor 9.1), XnXn
\citep{Adam:2019mby} and STARLight \citep{Klein:2016yzr}, respectively.
The data points are from STAR measurements \citep{Adam:2019mby},
while the shaded areas stand for the experimental uncertainty. \label{fig:Pt}}
\end{figure}

\begin{figure}
\centering{}\includegraphics[scale=0.4]{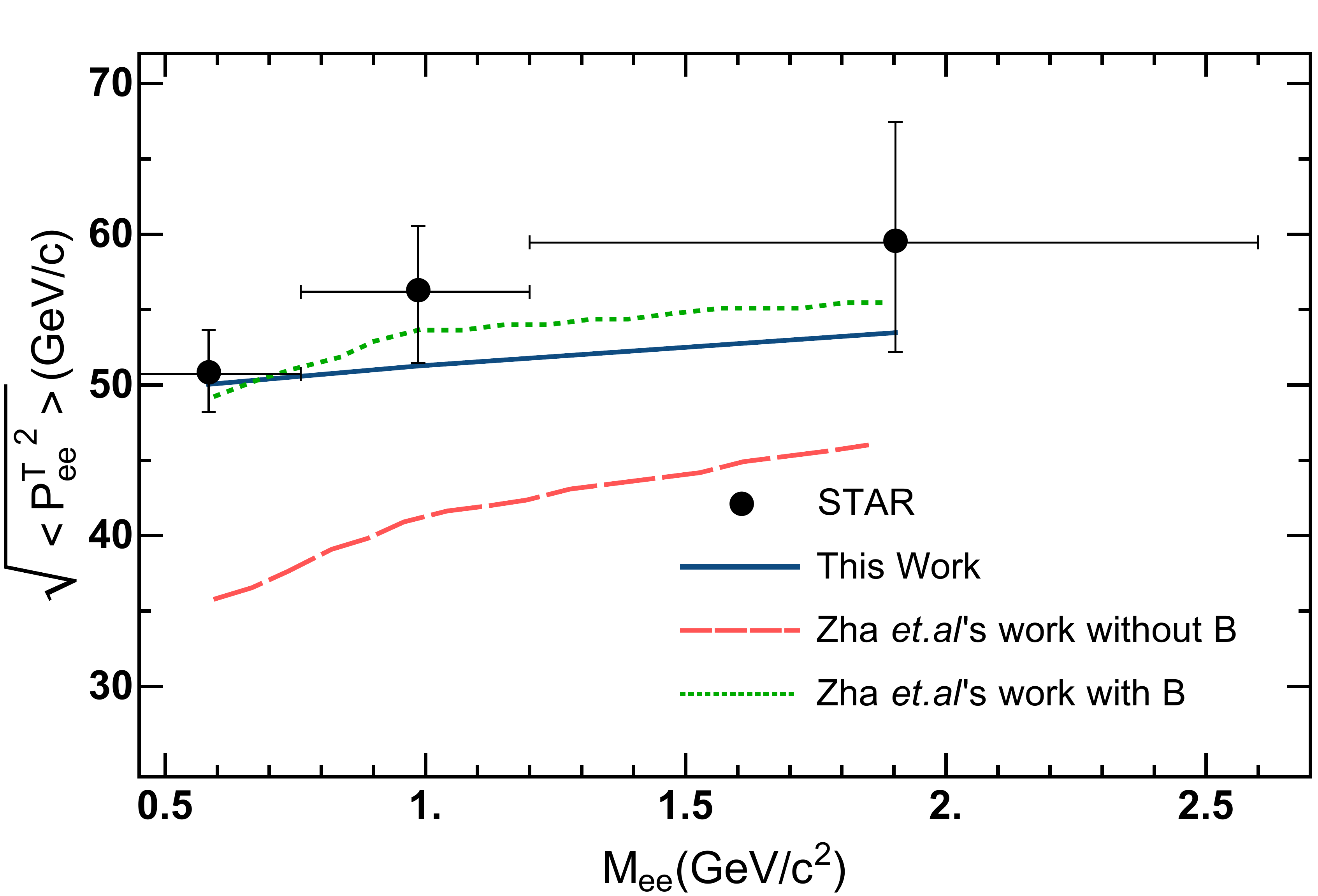}\caption{Average transverse momentum squares of lepton pairs as functions of
$M_{ee}$. The blue-solid line is our results using Eq. (\ref{eq:cs-photon-impact}).
The green-dotted lines and red-dashed are from Ref. \citep{Zha:2018ywo}
with and without contributions of magnetic fields, respectively. The
data points are from STAR measurements \citep{Adam:2019mby}. \label{fig:PtMee}}
\end{figure}

\begin{figure}
\includegraphics[scale=0.35]{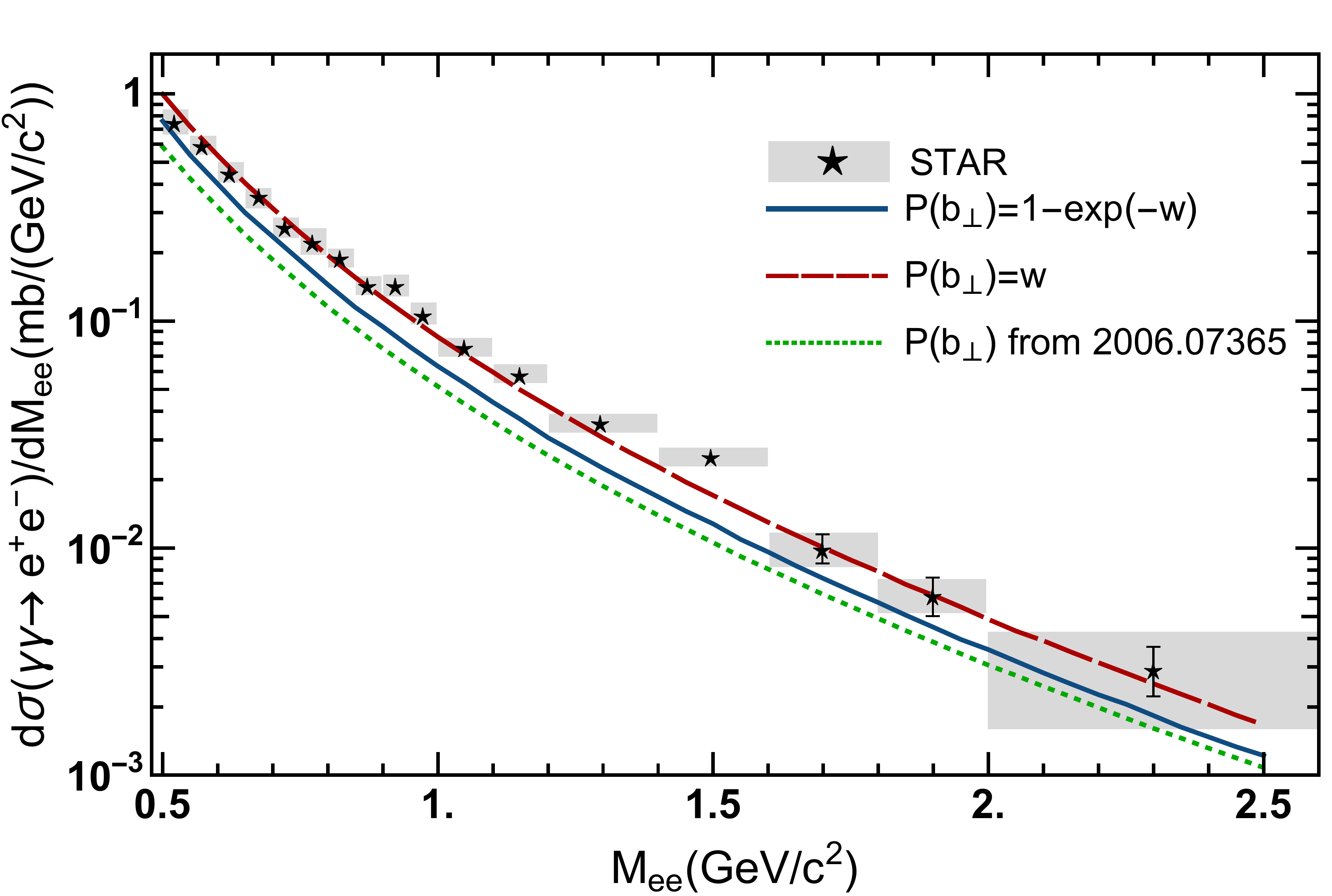}

\includegraphics[scale=0.34]{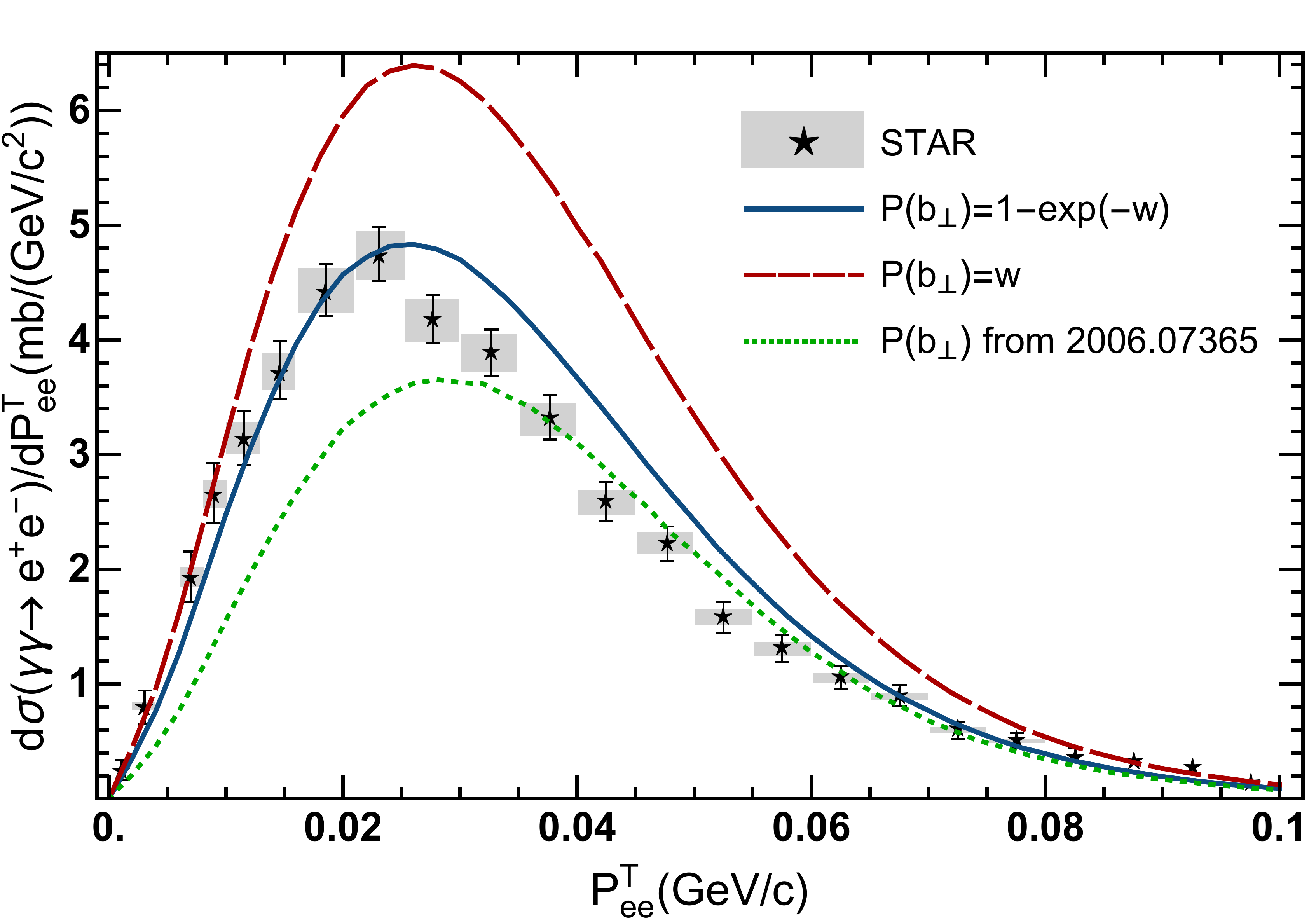}

\caption{Differential cross sections by using different $\mathcal{P}(b_{T}).$
The blue-solid, dark-red-dashed and green-dotted lines are the results
of using $\mathcal{P}(b_{T})=1-\exp(-w)$, $\mathcal{P}(b_{T})=w$,
and $\mathcal{P}(b_{T})$ of Ref. \citep{Brandenburg:2020ozx}, respectively.
\label{fig:DifferentP}}
\end{figure}

\section{Numerical results \label{sec:Numerical-results}}

In this section, we will present numerical results for the cross section
and compare them with STAR measurements at 200 GeV Au+Au collisions
at RHIC \citep{Adam:2018tdm,Adam:2019mby}. The cross section can
be expressed as a multi-dimension integral over independent variables
\begin{eqnarray}
\sigma & = & \frac{Z^{4}\alpha^{2}}{64\pi^{4}\gamma^{4}v^{2}}\int db_{T}d\phi_{b}\frac{d\Delta_{T}d\phi_{\Delta}}{(2\pi)^{2}}\frac{dp_{1T}d\phi_{p1}}{(2\pi)^{2}}\int dY_{ee}dP_{ee}^{T}d\phi_{ee}dM_{ee}d\eta_{k2}d\phi_{k2}\nonumber \\
 &  & \times e^{-i\mathbf{b}_{T}\cdot\boldsymbol{\Delta}_{T}}b_{T}\Delta_{T}p_{1T}\mathcal{P}^{2}(b_{T})\mathcal{J}\frac{F^{*}(-p_{1}^{\prime2})}{-p_{1}^{\prime2}}\frac{F(-p_{1}^{2})}{-p_{1}^{2}}\frac{F^{*}(-p_{2}^{\prime2})}{-p_{2}^{\prime2}}\frac{F(-p_{2}^{2})}{-p_{2}^{2}}\nonumber \\
 &  & \times\sum_{\textrm{spin of }l,\overline{l}}\left[u_{1\mu}u_{2\nu}L^{\mu\nu}(p_{1},p_{2};k_{1},k_{2})\right]\left[u_{1\sigma}u_{2\rho}L^{\sigma\rho*}(p_{1}^{\prime},p_{2}^{\prime};k_{1},k_{2})\right].\label{eq:cross-section-indep-var}
\end{eqnarray}
The derivation of the above formula as well as the explanation of
all variables are given in Appendix \ref{sec:Integral-variables}.
Here we have introduced a factor $\mathcal{P}^{2}(b_{T})$ which will
be explained shortly. The high-dimension integral of the cross section
(\ref{eq:cross-section-indep-var}) is a challenge in the numerical
calculation. In this paper we use the ZMCintegral package \citep{Wu_2020cpc_wzpw,Zhang_2020cpc_zw}
which has been applied to calculation of high dimensional integrals
in heavy ion collisions \citep{Zhang:2019uor}. 

For the numerical calculation we choose the nucleus form factor \citep{Klein:2016yzr,Klein:1999qj,Li:2019sin,Li:2019yzy}
as 
\begin{equation}
F(q)=\frac{4\pi\rho^{0}}{q^{3}A}[\sin(qR_{A})-qR_{A}\cos(qR_{A})]\frac{1}{a^{2}q^{2}+1},
\end{equation}
where $a=0.7$ fm, $R_{A}=1.2A^{1/3}$ fm is the nucleus radius with
$A$ being the number of nucleons, and $\rho^{0}=3A/(4\pi R_{A}^{3})$
is the nucleon number density to make $F(q=0)=1$ \citep{Klein:1999qj}.
One can also choose $R_{A}=1.1A^{1/3}\textrm{fm}$ as in Ref. \citep{Li:2019sin,Li:2019yzy}.

Another factor that should be taken into account is that two nuclei
may undergo mutual Coulomb excitation and emit neutrons \citep{Li:2019sin}.
To consider such effects, we need to introduce an extra factor $\mathcal{P}^{2}(b_{T})$
in the calculation of the cross section, where $\mathcal{P}(b_{T})$
is the probability of emitting a single neutron from an excited nucleus
and can be parametrized as \citep{Bertulani:1987tz},
\begin{equation}
\mathcal{P}(b_{T})=\sum_{N_{\gamma}=1}^{\infty}\frac{1}{N_{\gamma}!}w^{N_{\gamma}}e^{-w}=1-\exp(-w),\label{eq:P_b_perp}
\end{equation}
where $N_{\gamma}$ denotes the number of photons that can be absorbed
by a nucleus, and $w$ is defined by
\begin{equation}
w=5.45\times10^{-5}\frac{Z^{3}(A-Z)}{A^{2/3}b_{T}^{2}}.
\end{equation}
Note that the probability of emitting more than one neutron is highly
suppressed, so multiple neutron emission is usually neglected. The
formula (\ref{eq:P_b_perp}) has been widely used \citep{Bertulani:1987tz,Baur:1998ay,Pshenichnov:2001qd,Hencken:2004td,Baltz:2009jk,Brandenburg:2020ozx}.
Other choices are $\mathcal{P}(b_{T})\simeq w$ \citep{Hencken:2004td}
and $\mathcal{P}(b_{T})=w\exp(-w)$ \citep{Bertulani:1987tz,Baltz:1996as,Baur:1998ay,Hencken:2004td,Brandenburg:2020ozx}.
In this work, we choose (\ref{eq:P_b_perp}) in our numerical calculation. 

We follow STAR experiments \citep{Adam:2018tdm,Adam:2019mby} to choose
that the transverse momentum of the electron and that of the positron
is greater than 200 MeV but the transverse momentum of the electron-positron
pair is less than 100 MeV.

In Table \ref{tab:Total_cross_section_01-1}, we show the total cross
sections from STAR measurements and some theoretical models. The total
cross sections computed from both Eq. (\ref{eq:cs-photon-impact})
(under the approximation $G_{T}\approx1$) and gEPA in Eq. (\ref{eq:sigma-0})
are very close to the experimental results.

In Fig. \ref{fig:Mee}, we plot differential cross sections as functions
of the invariant mass of the lepton pair $M_{ee}$. The STAR data
\citep{Adam:2019mby} as well as the results of some models \citep{Zha:2018ywo,Klein:2016yzr}
(including our model) are presented. Our results using Eq. (\ref{eq:cs-photon-impact})
under the approximation $G_{T}\approx1$ (the blue-solid line) are
consistent with data. But our results give lower values than those
from gEPA.

The differential cross sections as functions of lepton pair's transverse
momenta are presented in Fig. \ref{fig:Pt}. We compare our results
with STAR data \citep{Adam:2019mby}, the results of Ref. \citep{Zha:2018ywo}
and STARLight \citep{Klein:2016yzr}. Our results using Eq. (\ref{eq:cs-photon-impact})
are in a good agreement with the data in the low transverse momentum
region, while they give a little larger values than the data for transverse
momenta larger than 0.02 GeV. The difference may come from possible
higher order corrections \citep{Sun:2020ygb,Zha:2021jhf}.

In Fig. \ref{fig:PtMee}, we show the average transverse momentum
squares of lepton pairs as functions of $M_{ee}$. Our results are
consistent with experimental data, so we do not need to consider extra
contributions from magnetic fields at the early stage. According to
Refs. \citep{Klein:2018fmp,Ringer:2019rfk,Klein:2020jom}, the Sudakov
form factor may play a role to the broadening of the transverse momentum.

That difference between our results and experimental data may come
from the choice of parameters. In Fig. \ref{fig:DifferentP}, we show
the results from different $\mathcal{P}(b_{T}).$ For the invariant
mass spectrum of the lepton pair, the result with $\mathcal{P}(b_{T})\simeq w$
\citep{Hencken:2004td} matches the data, but the transverse momentum
spectrum of the lepton pair is much higher than the data. The result
with $\mathcal{P}(b_{T})$ of Ref. \citep{Brandenburg:2020ozx} under-estimate
both invariant mass and transverse momentum spectra. These results
imply that we need to study the parameter dependence systematically,
which we leave for a future study.


\section{Summary \label{sec:Summary}}

A general form for the cross section of the lepton pair production
in ultra-peripheral collisions of heavy ions is derived. The wave
functions of two colliding nuclei moving in $\pm z$ direction are
assumed to be in the form of wave packets which allow a rigorous description
of ultra-peripheral collisions. A relative phase factor is introduced
into the wave function of one colliding nucleus that is displaced
by an impact parameter from the other colliding nucleus, through which
the transverse momentum and position are coupled. This leads to photon
distributions with dependence on transverse momentum and position.

The results of the generalized equivalent photon approximation can
be reproduced at the ultra-relativistic limit in our formalism. The
results of the TMD factorization formalism up to the Born approximation
can also be reproduced from our general form if light-cone coordinates
are used. It can be proved that the results of the generalized equivalent
photon approximation are consistent to the twist-2 results of the
TMD factorization formalism in the classical field approximation at
the Born level. We have also shown that the general form of the cross
section has already included high order corrections such as the ones
beyond the equivalent photon approximation or the ones from higher
twists in the classical field approximation at the Born level.

The numerical results for the differential cross sections with respect
to invariant mass and transverse momentum of the lepton pair are in
a good agreement with STAR data. One approximation we have made in
our results is that wave packets are assumed to be narrow so that
they are similar to plane waves. In the future this approximation
will be relaxed so that the effects from broader wave packets can
be studied. Other effects such as Sudakov factor as well as the dependence
on various parameters are expected to be investigated in the future.

\begin{acknowledgments}
We would like to thank Jian Zhou, Zhangbu Xu, Wangmei Zha and Zebo
Tang for helpful discussion. S.P. is supported by National Nature
Science Foundation of China (NSFC) under Grants No. 12075235. Q.W.
is supported in part by NSFC under Grant No. 11890713 (a sub-grant
of 11890710), and 11947301, and by the Strategic Priority Research
Program of Chinese Academy of Sciences under Grant No. XDB34030102.
\end{acknowledgments}

\appendix

\section{Derivation of impact-parameter dependent cross section \label{sec:Derive}}

In this appendix, we will derive the cross section which depends on
the impact parameter $\mathbf{b}_{T}$. We assume the wave functions
of two colliding nuclei moving in $\pm z$ direction (see Fig. \ref{fig:collision})
can be written in the form of wave packets following the argument
of Ref. \citep{Peskin:1995ev},
\begin{eqnarray}
\left|A_{1}\right\rangle  & = & \int\frac{d^{3}P_{1}}{(2\pi)^{3}\sqrt{2E_{P1}}}\phi(P_{1})e^{ib\cdot P_{1}}\left|P_{1}\right\rangle ,\nonumber \\
\left|A_{2}\right\rangle  & = & \int\frac{d^{3}P_{2}}{(2\pi)^{3}\sqrt{2E_{P2}}}\phi(P_{2})\left|P_{2}\right\rangle ,\label{eq:wave_function_01}
\end{eqnarray}
where $P_{1}=(E_{P1},\mathbf{P}_{1})$ and $P_{2}=(E_{P2},\mathbf{P}_{2})$
with $E_{P1}=\sqrt{\mathbf{P}_{1}^{2}+M_{1}^{2}}$ and $E_{P2}=\sqrt{\mathbf{P}_{2}^{2}+M_{2}^{2}}$
are on-shell momenta of two nuclei with masses $M_{1}$ and $M_{2}$,
$\phi(P_{1})\equiv\phi(\mathbf{P}_{1}-\mathbf{P}_{A1})$ and $\phi(P_{2})\equiv\phi(\mathbf{P}_{2}-\mathbf{P}_{A2})$
denote the amplitudes of momentum states which are centered at the
nuclear momenta $\mathbf{P}_{A1}=(0,0,P_{A1}^{z})$ and $\mathbf{P}_{A2}=(0,0,-P_{A1}^{z})$
respectively. They satisfy the normalization condition $(2\pi)^{-3}\int d^{3}k|\phi(k)|^{2}=1$.
Here we have introduced the impact parameter $b^{\mu}=(0,\mathbf{b}_{T},0)$
into the state of $A_{1}$, which is the transverse distance between
the centers of two nuclei. 

The initial state of nuclei can be written as 
\begin{equation}
\left|A_{1}A_{2}\right\rangle _{\mathrm{in}}=\int\frac{d^{3}P_{1}}{(2\pi)^{3}}\frac{d^{3}P_{2}}{(2\pi)^{3}}\frac{\phi(P_{1})\phi(P_{2})e^{ib_{T}\cdot P_{1}}}{\sqrt{2E_{P1}}\sqrt{2E_{P2}}}\left|P_{1}P_{2}\right\rangle _{\mathrm{in}},\label{eq:nuclear-state}
\end{equation}
while the final state is assumed to be in momentum states instead
of wave packets
\begin{equation}
\left|l,\overline{l},\sum_{f}X_{f}\right\rangle \equiv\left|k_{1},k_{2},\sum_{f}K_{f}\right\rangle _{\mathrm{out}},
\end{equation}
where the indices 'in' and 'out' stand for the in- and out- states
at $t\rightarrow\pm\infty$, respectively, and $X_{f}$ with momenta
$K_{f}$ denote all other particles that are produced in collisions.

The cross section reads 
\begin{eqnarray}
\sigma & = & \int d^{2}\mathbf{b}_{T}\sum_{\{f\}}\int\frac{d^{3}k_{1}}{(2\pi)^{3}2E_{k_{1}}}\frac{d^{3}k_{2}}{(2\pi)^{3}2E_{k_{2}}}\prod_{f}\frac{d^{3}K_{f}}{(2\pi)^{3}2E_{f}}\nonumber \\
 &  & \times\left|_{\mathrm{out}}\left\langle k_{1},k_{2},\right.\sum_{f}K_{f}\left|A_{1}A_{2}\right\rangle _{\mathrm{in}}\right|^{2}.
\end{eqnarray}
Using Eqs. (\ref{eq:nuclear-state},\ref{eq:T_matrix_01}) and completing
the integration over $P_{1}^{\prime z}$ and $P_{2}^{\prime z}$ in
the above formula, we obtain 
\begin{eqnarray}
\sigma & = & \int d^{2}\mathbf{b}_{T}\sum_{\{f\}}\int\frac{d^{3}k_{1}}{(2\pi)^{3}2E_{k1}}\frac{d^{3}k_{2}}{(2\pi)^{3}2E_{k2}}\prod_{f}\frac{d^{3}K_{f}}{(2\pi)^{3}2E_{Kf}}\nonumber \\
 &  & \times\int\frac{d^{3}P_{1}}{(2\pi)^{3}\sqrt{2E_{P1}}}\frac{d^{3}P_{2}}{(2\pi)^{3}\sqrt{2E_{P2}}}\frac{d^{3}P_{1}^{\prime}}{(2\pi)^{3}\sqrt{2E_{P1^{\prime}}}}\frac{d^{3}P_{2}^{\prime}}{(2\pi)^{3}\sqrt{2E_{P2^{\prime}}}}\nonumber \\
 &  & \times\phi(P_{1})\phi(P_{2})\phi^{*}(P_{1}^{\prime})\phi^{*}(P_{2}^{\prime})e^{-i\mathbf{b}_{T}\cdot\boldsymbol{\Delta}_{1T}}\nonumber \\
 &  & \times(2\pi)^{4}\delta^{(4)}\left(P_{1}+P_{2}-k_{1}-k_{2}-\sum_{f}K_{f}\right)(2\pi)^{4}\delta^{(4)}\left(P_{1}+P_{2}-P_{1}^{\prime}-P_{2}^{\prime}\right)\nonumber \\
 &  & \times\sum_{\textrm{spin of }l,\overline{l}}\mathcal{M}_{P_{1}+P_{2}\rightarrow k_{1}+k_{2}+\sum_{f}K_{f}}\mathcal{M}_{P_{1}^{\prime}+P_{2}^{\prime}\rightarrow k_{1}+k_{2}+\sum_{f}K_{f}}^{*},\label{eq:cross-bt}
\end{eqnarray}
where $\boldsymbol{\Delta}_{1T}\equiv\mathbf{P}_{1T}^{\prime}-\mathbf{P}_{1T}$,
$\mathcal{M}$ denotes the invariant amplitude defined through the
T-matrix element. By rewriting the delta function for transverse momenta
as 
\begin{equation}
\delta^{(2)}\left(\mathbf{P}_{1T}+\mathbf{P}_{2T}-\mathbf{P}_{1T}^{\prime}-\mathbf{P}_{2T}^{\prime}\right)=\int\frac{d^{2}\mathbf{b}_{2T}}{(2\pi)^{2}}\exp\left[i\mathbf{b}_{2T}\cdot\left(\mathbf{P}_{1T}+\mathbf{P}_{2T}-\mathbf{P}_{1T}^{\prime}-\mathbf{P}_{2T}^{\prime}\right)\right],
\end{equation}
and adding an integral 
\begin{equation}
\int d^{2}\mathbf{b}_{1T}\delta^{(2)}\left(\mathbf{b}_{T}-\mathbf{b}_{1T}+\mathbf{b}_{2T}\right)=1,
\end{equation}
The cross section (\ref{eq:cross-bt}) can now be put into the form
\begin{eqnarray}
\sigma & = & \int d^{2}\mathbf{b}_{T}d^{2}\mathbf{b}_{1T}d^{2}\mathbf{b}_{2T}\sum_{\{f\}}\int\frac{d^{3}k_{1}}{(2\pi)^{3}2E_{k1}}\frac{d^{3}k_{2}}{(2\pi)^{3}2E_{k2}}\prod_{f}\frac{d^{3}K_{f}}{(2\pi)^{3}2E_{Kf}}\nonumber \\
 &  & \times\int\frac{d^{3}P_{1}}{(2\pi)^{3}\sqrt{2E_{P1}}}\frac{d^{3}P_{2}}{(2\pi)^{3}\sqrt{2E_{P2}}}\frac{d^{3}P_{1}^{\prime}}{(2\pi)^{3}\sqrt{2E_{P1^{\prime}}}}\frac{d^{3}P_{2}^{\prime}}{(2\pi)^{3}\sqrt{2E_{P2^{\prime}}}}\nonumber \\
 &  & \times\phi(P_{1})\phi(P_{2})\phi^{*}(P_{1}^{\prime})\phi^{*}(P_{2}^{\prime})\nonumber \\
 &  & \times e^{-i\mathbf{b}_{1T}\cdot\Delta_{1T}}e^{-i\mathbf{b}_{2T}\cdot\Delta_{2T}}\delta^{(2)}\left(\mathbf{b}_{T}-\mathbf{b}_{1T}+\mathbf{b}_{2T}\right)\nonumber \\
 &  & \times(2\pi)^{4}\delta^{(4)}\left(P_{1}+P_{2}-k_{1}-k_{2}-\sum_{f}K_{f}\right)\nonumber \\
 &  & \times(2\pi)^{2}\delta(P_{1}^{\prime z}+P_{2}^{\prime z}-P_{1}^{z}-P_{2}^{z})\delta(E_{P1^{\prime}}+E_{P2^{\prime}}-E_{P1}-E_{P2})\nonumber \\
 &  & \times\sum_{\textrm{spin of }l,\overline{l}}\mathcal{M}_{P_{1}+P_{2}\rightarrow k_{1}+k_{2}+\sum_{f}K_{f}}\mathcal{M}_{P_{1}^{\prime}+P_{2}^{\prime}\rightarrow k_{1}+k_{2}+\sum_{f}K_{f}}^{*},\label{eq:cross-bt-1}
\end{eqnarray}
where we have used $\boldsymbol{\Delta}_{2T}\equiv\mathbf{P}_{2T}^{\prime}-\mathbf{P}_{2T}$.

Now we look at the momentum amplitudes $\phi(P_{1})$ and $\phi(P_{2})$
in nuclear wave functions as wave packets. Normally the total cross
section does not depend on the form of $\phi(P_{1})$ and $\phi(P_{2})$
\citep{Peskin:1995ev}. A natural choice for $\phi(P_{1})$ and $\phi(P_{2})$
is that they can be factorized as $\phi(P_{i})\simeq\phi_{z}(P_{i}^{z})\phi_{T}(\mathbf{P}_{iT})$
with $i=1,2$, where $\phi_{z}(P_{i}^{z})$ are distributions of longitudinal
momenta that $P_{1}^{z}$ and $P_{2}^{z}$ are centered at $P_{A1}^{z}$
and $-P_{A1}^{z}$ respectively, and $\phi_{T}(\mathbf{P}_{iT})$
are distributions of transverse momenta that $\mathbf{P}_{iT}$ are
centered at zero. As a simple ansatz we assume the longitudinal part
takes the form 
\begin{eqnarray}
 &  & \phi_{z}(P_{1}^{z})\phi_{z}(P_{2}^{z})\phi_{z}^{*}(P_{1}^{\prime z})\phi_{z}^{*}(P_{2}^{\prime z})\nonumber \\
 & \approx & (2\pi)^{2}\delta\left[(P_{1}^{z}+P_{1}^{\prime z})/2-P_{A1}^{z}\right]\delta\left[(P_{2}^{z}+P_{2}^{\prime z})/2+P_{A1}^{z}\right]\nonumber \\
 &  & \times G\left[\frac{1}{4}(P_{1}^{z}-P_{1}^{\prime z})^{2}\right]G\left[\frac{1}{4}(P_{2}^{z}-P_{2}^{\prime z})^{2}\right],\label{eq:wave_ansatz}
\end{eqnarray}
where $G(x^{2})$ is a positive function with $G(0)=1$ and decreases
rapidly with growing $x^{2}$, and the factor $1/4$ inside $G$-functions
is by convention. So we can carry out the integrals over $P_{1}^{z}$
and $P_{2}^{z}$ to remove two delta-functions and setting $P_{1}^{z}=2P_{A1}^{z}-P_{1}^{\prime z}$
and $P_{2}^{z}=-2P_{A1}^{z}-P_{2}^{\prime z}$ in the integrand. Then
we complete the integrals over $P_{1}^{\prime z}$ and $P_{2}^{\prime z}$
to remove two delta-functions for longitudinal momenta and energies
\begin{eqnarray}
 &  & \int dP_{1z}^{\prime}dP_{2z}^{\prime}\delta(P_{1}^{\prime z}+P_{2}^{\prime z}-P_{1}^{z}-P_{2}^{z})\delta(E_{P1^{\prime}}+E_{P2^{\prime}}-E_{P1}-E_{P2})\nonumber \\
 & = & \frac{1}{2}\int dP_{1z}^{\prime}dP_{2z}^{\prime}\delta(P_{1}^{\prime z}+P_{2}^{\prime z})\delta(E_{P1^{\prime}}+E_{P2^{\prime}}-E_{P1}-E_{P2})\nonumber \\
 & = & \frac{1}{2\left|P_{1}^{\prime z}/E_{P1^{\prime}}+P_{1}^{\prime z}/E_{P2^{\prime}}+P_{1}^{z}/E_{P1}+P_{1}^{z}/E_{P2}\right|}\equiv\frac{1}{8v},\label{eq:p1z-p2z-prime}
\end{eqnarray}
where we have used $P_{1}^{z}=2P_{A1}^{z}-P_{1}^{\prime z}$ and $P_{2}^{z}=-2P_{A1}^{z}-P_{2}^{\prime z}$,
and $P_{1}^{\prime z}=-P_{2}^{\prime z}$ are solved as functions
of transverse momenta from the energy conservation. Furthermore the
condition $P_{1}^{\prime z}=-P_{2}^{\prime z}$ leads to $P_{1}^{z}=-P_{2}^{z}=2P_{A1}^{z}-P_{1}^{\prime z}$.
With these conditions, two $G$-functions in Eq. (\ref{eq:wave_ansatz})
give $G^{2}\left[(P_{1}^{\prime z}-P_{A1}^{z})^{2}\right]$ as a function
of transverse momenta.

Completing the integrals over $P_{1}^{z}$, $P_{2}^{z}$, $P_{1}^{\prime z}$
and $P_{2}^{\prime z}$ in Eq. (\ref{eq:cross-bt-1}) by using (\ref{eq:wave_ansatz})
and (\ref{eq:p1z-p2z-prime}), we obtain Eq. (\ref{eq:cross-section-4})
for the cross section.

Now we deal with the invariant amplitude $\mathcal{M}$ through the
matrix element of the operator $\widehat{T}$ (T-matrix element) as
\begin{equation}
_{\mathrm{out}}\left\langle k_{1},k_{2},\right.\sum_{f}K_{f}\left|P_{1}P_{2}\right\rangle _{\mathrm{in}}=\left\langle k_{1},k_{2},\sum_{f}K_{f}|(1+i\widehat{T})|P_{1}P_{2}\right\rangle 
\end{equation}
with the T-matrix element being parametrized as 
\begin{eqnarray}
\left\langle k_{1},k_{2},\sum_{f}K_{f}|i\widehat{T}|P_{1}P_{2}\right\rangle  & = & (2\pi)^{4}\delta^{(4)}\left(P_{1}+P_{2}-k_{1}-k_{2}-\sum_{f}K_{f}\right)\nonumber \\
 &  & \times i\mathcal{M}_{P_{1}+P_{2}\rightarrow k_{1}+k_{2}+\sum_{f}K_{f}},\label{eq:T_matrix_01}
\end{eqnarray}
where $\mathcal{M}$ denotes the invariant amplitude of the process.
By definition, the T-matrix element can be written in the form
\begin{eqnarray}
\left\langle k_{1},k_{2},\sum_{f}K_{f}|i\widehat{T}|P_{1}P_{2}\right\rangle  & = & -e^{2}\int d^{4}x_{1}d^{4}x_{2}\nonumber \\
 &  & \times\left\langle k_{1}k_{2}\right|\mathcal{T}\left[\overline{\psi}(x_{1})\gamma^{\mu}\psi(x_{1})\overline{\psi}_{2}(x_{2})\gamma^{\nu}\psi_{2}(x_{2})\right]\left|0\right\rangle \nonumber \\
 &  & \times\left\langle \sum_{f}K_{f}|\mathcal{T}\left[A_{\mu}(x_{1})A_{\nu}(x_{2})\right]|P_{1}P_{2}\right\rangle ,\label{eq:t-element}
\end{eqnarray}
where $\mathcal{T}$ denotes the time-ordered operator. The lepton
part is evaluated as 
\begin{eqnarray}
 &  & \left\langle k_{1}k_{2}\right|\mathcal{T}\left[\overline{\psi}(x_{1})\gamma^{\mu}\psi(x_{1})\overline{\psi}_{2}(x_{2})\gamma^{\nu}\psi_{2}(x_{2})\right]\left|0\right\rangle \nonumber \\
 & = & i\int\frac{d^{4}q}{(2\pi)^{4}}\overline{u}(k_{1})\gamma^{\mu}\frac{\gamma\cdot q+m}{q^{2}-m^{2}+i\varepsilon}\gamma^{\nu}v(k_{2})e^{ik_{1}\cdot x_{1}+ik_{2}\cdot x_{2}-iq\cdot(x_{1}-x_{2})}\nonumber \\
 &  & +i\int\frac{d^{4}q}{(2\pi)^{4}}\overline{u}(k_{1})\gamma^{\nu}\frac{\gamma\cdot q+m}{q^{2}-m^{2}+i\varepsilon}\gamma^{\mu}v(k_{2})e^{ik_{1}\cdot x_{2}+ik_{2}\cdot x_{1}+iq\cdot(x_{1}-x_{2})}.\label{eq:lepton-part}
\end{eqnarray}
We make an approximation that the time-ordered operator in the EM
part can be removed so that it can be put into the form 
\begin{eqnarray}
\left\langle \sum_{f}K_{f}|\mathcal{T}\left[A_{\mu}(x_{1})A_{\nu}(x_{2})\right]|P_{1}P_{2}\right\rangle  & \sim & \left\langle \sum_{f}K_{f}|A_{\mu}(x_{1})A_{\nu}(x_{2})|P_{1}P_{2}\right\rangle \nonumber \\
 & = & \int\frac{d^{4}p_{1}}{(2\pi)^{4}}\int\frac{d^{4}p_{2}}{(2\pi)^{4}}\exp\left(-ip_{1}\cdot x_{1}-ip_{2}\cdot x_{2}\right)\nonumber \\
 &  & \times\left\langle \sum_{f}K_{f}|A_{\mu}(p_{1})A_{\nu}(p_{2})|P_{1}P_{2}\right\rangle ,\label{eq:photon-part}
\end{eqnarray}
where we have performed Fourier transformation for the EM fields.
Using Eqs. (\ref{eq:lepton-part},\ref{eq:photon-part}) and completing
the integrals over $x_{1}$ and $x_{2}$ in Eq. (\ref{eq:t-element}),
we obtain 
\begin{eqnarray}
\left\langle k_{1},k_{2},\sum_{f}K_{f}|i\widehat{T}|P_{1}P_{2}\right\rangle  & = & (2\pi)^{4}\delta^{(4)}\left(P_{1}+P_{2}-k_{1}-k_{2}-\sum_{f}K_{f}\right)\nonumber \\
 &  & \times i\mathcal{M}_{P_{1}+P_{2}\rightarrow k_{1}+k_{2}+\sum_{f}K_{f}}\nonumber \\
\mathcal{M}_{P_{1}+P_{2}\rightarrow k_{1}+k_{2}+\sum_{f}K_{f}} & = & -e^{2}\int\frac{d^{4}p_{1}}{(2\pi)^{4}}\int\frac{d^{4}p_{2}}{(2\pi)^{4}}\left\langle \sum_{f}K_{f}|A_{\mu}(p_{1})A_{\nu}(p_{2})|P_{1}P_{2}\right\rangle \nonumber \\
 &  & \times\left[\overline{u}(k_{1})\gamma^{\mu}\frac{\gamma\cdot(k_{1}-p_{1})+m}{(k_{1}-p_{1})^{2}-m^{2}+i\varepsilon}\gamma^{\nu}v(k_{2})\right.\nonumber \\
 &  & \left.+\overline{u}(k_{1})\gamma^{\nu}\frac{\gamma\cdot(p_{1}-k_{2})+m}{(p_{1}-k_{2})^{2}-m^{2}+i\varepsilon}\gamma^{\mu}v(k_{2})\right].\label{eq:it-matrix}
\end{eqnarray}
Following the same procedure, we obtain 
\begin{eqnarray}
\left\langle P_{1}^{\prime}P_{2}^{\prime}|(-i\widehat{T}^{\dagger})|k_{1},k_{2},\sum_{f}K_{f}\right\rangle  & = & (2\pi)^{4}\delta^{(4)}\left(P_{1}^{\prime}+P_{2}^{\prime}-k_{1}-k_{2}-\sum_{f}K_{f}\right)\nonumber \\
 &  & \times(-i)\mathcal{M}_{P_{1}^{\prime}+P_{2}^{\prime}\rightarrow k_{1}+k_{2}+\sum_{f}K_{f}}^{*}\nonumber \\
\mathcal{M}_{P_{1}^{\prime}+P_{2}^{\prime}\rightarrow k_{1}+k_{2}+\sum_{f}K_{f}}^{*} & = & -e^{2}\int\frac{d^{4}p_{1}^{\prime}}{(2\pi)^{4}}\int\frac{d^{4}p_{2}^{\prime}}{(2\pi)^{4}}\left\langle P_{1}^{\prime}P_{2}^{\prime}|A_{\nu}^{\dagger}(p_{2}^{\prime})A_{\mu}^{\dagger}(p_{1}^{\prime})|\sum_{f}K_{f}\right\rangle \nonumber \\
 &  & \times\left[\overline{v}(k_{2})\gamma^{\nu}\frac{\gamma\cdot(k_{1}-p_{1}^{\prime})+m}{(k_{1}-p_{1}^{\prime})^{2}-m^{2}+i\varepsilon}\gamma^{\mu}u(k_{1})\right.\nonumber \\
 &  & \left.+\overline{v}(k_{2})\gamma^{\mu}\frac{\gamma\cdot(p_{1}^{\prime}-k_{2})+m}{(p_{1}^{\prime}-k_{2})^{2}-m^{2}+i\varepsilon}\gamma^{\nu}u(k_{1})\right].\label{eq:it-star-matrix}
\end{eqnarray}
Note that in Eqs. (\ref{eq:it-matrix},\ref{eq:it-star-matrix}) we
have suppressed the spin indices of lepton spinors. 

Using Eqs. (\ref{eq:it-matrix},\ref{eq:it-star-matrix}) in Eq. (\ref{eq:cross-section-4}),
the cross section is put into the form 
\begin{eqnarray}
\sigma & = & \frac{1}{8(2\pi)^{8}}\int d^{2}\mathbf{b}_{T}d^{2}\mathbf{b}_{1T}d^{2}\mathbf{b}_{2T}\int\frac{d^{3}k_{1}}{(2\pi)^{3}2E_{k1}}\frac{d^{3}k_{2}}{(2\pi)^{3}2E_{k2}}\nonumber \\
 &  & \times\int d^{2}\mathbf{P}_{1T}d^{2}\mathbf{P}_{2T}d^{2}\mathbf{P}_{1T}^{\prime}d^{2}\mathbf{P}_{2T}^{\prime}\frac{1}{v\sqrt{E_{P1}E_{P2}E_{P1^{\prime}}E_{P2^{\prime}}}}\nonumber \\
 &  & \times\phi_{T}(\mathbf{P}_{1T})\phi_{T}(\mathbf{P}_{2T})\phi_{T}^{*}(\mathbf{P}_{1T}^{\prime})\phi_{T}^{*}(\mathbf{P}_{2T}^{\prime})G^{2}\left[(P_{1}^{\prime z}-P_{A1}^{z})^{2}\right]\nonumber \\
 &  & \times e^{-i\mathbf{b}_{1T}\cdot\boldsymbol{\Delta}_{1T}}e^{-i\mathbf{b}_{2T}\cdot\boldsymbol{\Delta}_{2T}}\delta^{(2)}\left(\mathbf{b}_{T}-\mathbf{b}_{1T}+\mathbf{b}_{2T}\right)\nonumber \\
 &  & \times\int\frac{d^{4}p_{1}}{(2\pi)^{4}}\frac{d^{4}p_{2}}{(2\pi)^{4}}\frac{d^{4}p_{1}^{\prime}}{(2\pi)^{4}}\frac{d^{4}p_{2}^{\prime}}{(2\pi)^{4}}\nonumber \\
 &  & \times\left\langle P_{1}^{\prime}\right|A_{\sigma}^{\dagger}(p_{1}^{\prime})A_{\mu}(p_{1})\left|P_{1}\right\rangle \left\langle P_{2}^{\prime}\right|A_{\rho}^{\dagger}(p_{2}^{\prime})A_{\nu}(p_{2})\left|P_{2}\right\rangle \nonumber \\
 &  & \times L^{\mu\nu;\sigma\rho}\left(p_{1},p_{2};p_{1}^{\prime},p_{2}^{\prime};k_{1},k_{2}\right)\nonumber \\
 &  & \times(2\pi)^{4}\delta^{(4)}\left(p_{1}+p_{2}-k_{1}-k_{2}\right),\label{eq:cross-section-2}
\end{eqnarray}
where we have used the identity 
\begin{equation}
\sum_{\{f\}}\int\prod_{f}\frac{d^{3}K_{f}}{(2\pi)^{3}2E_{Kf}}\left|\sum_{f}K_{f}\right\rangle \left\langle \sum_{f}K_{f}\right|=1,
\end{equation}
and the relation 
\begin{eqnarray}
\sum_{f}K_{f} & = & P_{1}+P_{2}-p_{1}-p_{2}\nonumber \\
 & = & P_{1}^{\prime}+P_{2}^{\prime}-p_{1}^{\prime}-p_{2}^{\prime}.
\end{eqnarray}
We have also used the lepton part as %
\begin{eqnarray}
 &  & L^{\mu\nu;\sigma\rho}\left(p_{1},p_{2};p_{1}^{\prime},p_{2}^{\prime};k_{1},k_{2}\right)\nonumber \\
 & = & e^{4}\sum_{\textrm{spin of }l,\overline{l}}L^{\mu\nu}(p_{1},p_{2};k_{1},k_{2})L^{\sigma\rho*}(p_{1}^{\prime},p_{2}^{\prime};k_{1},k_{2})\nonumber \\
 & = & e^{4}\textrm{Tr }\left\{ (\gamma\cdot k_{1}+m)\left[\gamma^{\mu}\frac{\gamma\cdot(k_{1}-p_{1})+m}{(k_{1}-p_{1})^{2}-m^{2}+i\varepsilon}\gamma^{\nu}+\gamma^{\nu}\frac{\gamma\cdot(p_{1}-k_{2})+m}{(p_{1}-k_{2})^{2}-m^{2}+i\varepsilon}\gamma^{\mu}\right]\right.\nonumber \\
 &  & \left.\times(\gamma\cdot k_{2}-m)\left[\gamma^{\rho}\frac{\gamma\cdot(k_{1}-p_{1}^{\prime})+m}{(k_{1}-p_{1}^{\prime})^{2}-m^{2}+i\varepsilon}\gamma^{\sigma}+\gamma^{\sigma}\frac{\gamma\cdot(p_{1}^{\prime}-k_{2})+m}{(p_{1}^{\prime}-k_{2})^{2}-m^{2}+i\varepsilon}\gamma^{\rho}\right]\right\} ,\label{eq:l-mu-nu-1}
\end{eqnarray}
where $L^{\mu\nu}$ and $L^{\sigma\rho*}$ are defined by 
\begin{eqnarray}
L^{\mu\nu}(p_{1},p_{2};k_{1},k_{2}) & = & \overline{u}(k_{1})\left[\gamma^{\mu}\frac{\gamma\cdot(k_{1}-p_{1})+m}{(k_{1}-p_{1})^{2}-m^{2}+i\varepsilon}\gamma^{\nu}\right.\nonumber \\
 &  & \left.+\gamma^{\nu}\frac{\gamma\cdot(p_{1}-k_{2})+m}{(p_{1}-k_{2})^{2}-m^{2}+i\varepsilon}\gamma^{\mu}\right]v(k_{2}),\nonumber \\
L^{\sigma\rho*}(p_{1}^{\prime},p_{2}^{\prime};k_{1},k_{2}) & = & \overline{v}(k_{2})\left[\gamma^{\rho}\frac{\gamma\cdot(k_{1}-p_{1}^{\prime})+m}{(k_{1}-p_{1}^{\prime})^{2}-m^{2}+i\varepsilon}\gamma^{\sigma}\right.\nonumber \\
 &  & \left.+\gamma^{\sigma}\frac{\gamma\cdot(p_{1}^{\prime}-k_{2})+m}{(p_{1}^{\prime}-k_{2})^{2}-m^{2}+i\varepsilon}\gamma^{\rho}\right]u(k_{1}).\label{eq:l-mu-nu-2}
\end{eqnarray}
A useful property of $L^{\mu\nu}(p_{1},p_{2},k_{1},k_{2})$ is the
following identity \citep{Vidovic:1992ik}
\begin{equation}
p_{1\mu}L^{\mu\nu}=p_{2\nu}L^{\mu\nu}=0.\label{eq:relation_Gamma_01}
\end{equation}
In (\ref{eq:cross-section-2}) we have also reduced the photon matrix
element as 
\begin{eqnarray}
 &  & \left\langle P_{1}^{\prime}P_{2}^{\prime}\right|A_{\rho}^{\dagger}(p_{2}^{\prime})A_{\sigma}^{\dagger}(p_{1}^{\prime})A_{\mu}(p_{1})A_{\nu}(p_{2})\left|P_{1}P_{2}\right\rangle \nonumber \\
 & \sim & \frac{1}{4}\left\langle P_{1}^{\prime}\right|A_{\sigma}^{\dagger}(p_{1}^{\prime})A_{\mu}(p_{1})\left|P_{1}\right\rangle \left\langle P_{2}^{\prime}\right|A_{\rho}^{\dagger}(p_{2}^{\prime})A_{\nu}(p_{2})\left|P_{2}\right\rangle \nonumber \\
 &  & +\frac{1}{4}\left\langle P_{1}^{\prime}\right|A_{\rho}^{\dagger}(p_{2}^{\prime})A_{\mu}(p_{1})\left|P_{1}\right\rangle \left\langle P_{2}^{\prime}\right|A_{\sigma}^{\dagger}(p_{1}^{\prime})A_{\nu}(p_{2})\left|P_{2}\right\rangle \nonumber \\
 &  & +\frac{1}{4}\left\langle P_{1}^{\prime}\right|A_{\sigma}^{\dagger}(p_{1}^{\prime})A_{\nu}(p_{2})\left|P_{1}\right\rangle \left\langle P_{2}^{\prime}\right|A_{\rho}^{\dagger}(p_{2}^{\prime})A_{\mu}(p_{1})\left|P_{2}\right\rangle \nonumber \\
 &  & +\frac{1}{4}\left\langle P_{1}^{\prime}\right|A_{\rho}^{\dagger}(p_{2}^{\prime})A_{\nu}(p_{2})\left|P_{1}\right\rangle \left\langle P_{2}^{\prime}\right|A_{\sigma}^{\dagger}(p_{1}^{\prime})A_{\mu}(p_{1})\left|P_{2}\right\rangle +\cdots,\nonumber \\
 & \sim & \left\langle P_{1}^{\prime}\right|A_{\sigma}^{\dagger}(p_{1}^{\prime})A_{\mu}(p_{1})\left|P_{1}\right\rangle \left\langle P_{2}^{\prime}\right|A_{\rho}^{\dagger}(p_{2}^{\prime})A_{\nu}(p_{2})\left|P_{2}\right\rangle .\label{eq:photon-me-4}
\end{eqnarray}
This is because each term in the second approximate equality can be
proved to be identical after change of variables, giving a symmetry
factor 4. So one can just take the first term and multiply it by 4
to obtain the photon matrix element. We can rewrite the integrals
over $p_{1}$, $p_{2}$, $p_{1}^{\prime}$ and $p_{2}^{\prime}$ in
(\ref{eq:cross-section-2}) as 
\begin{eqnarray}
I & = & \int\frac{d^{4}p_{1}}{(2\pi)^{4}}\frac{d^{4}p_{2}}{(2\pi)^{4}}\frac{d^{4}p_{1}^{\prime}}{(2\pi)^{4}}\frac{d^{4}p_{2}^{\prime}}{(2\pi)^{4}}L^{\mu\nu;\sigma\rho}(p_{1},p_{2};p_{1}^{\prime},p_{2}^{\prime};k_{1},k_{2})\nonumber \\
 &  & \times\left\langle P_{1}^{\prime}\right|A_{\sigma}^{\dagger}(p_{1}^{\prime})A_{\mu}(p_{1})\left|P_{1}\right\rangle \left\langle P_{2}^{\prime}\right|A_{\rho}^{\dagger}(p_{2}^{\prime})A_{\nu}(p_{2})\left|P_{2}\right\rangle \nonumber \\
 & = & \int\frac{d^{4}p_{1}}{(2\pi)^{4}}\frac{d^{4}p_{2}}{(2\pi)^{4}}\int d^{4}y_{1}d^{4}y_{2}\exp\left(ip_{1}\cdot y_{1}+ip_{2}\cdot y_{2}\right)\nonumber \\
 &  & \times L^{\mu\nu;\sigma\rho}(p_{1},p_{2};p_{1}-P_{1}+P_{1}^{\prime},p_{2}-P_{2}+P_{2}^{\prime};k_{1},k_{2})\nonumber \\
 &  & \times\left\langle P_{1}^{\prime}\right|A_{\sigma}^{\dagger}(0)A_{\mu}\left(y_{1}\right)\left|P_{1}\right\rangle \left\langle P_{2}^{\prime}\right|A_{\rho}^{\dagger}(0)A_{\nu}\left(y_{2}\right)\left|P_{2}\right\rangle .\label{eq:photon-me-x-space}
\end{eqnarray}
In deriving the above formula, we have converted all photon fields
in momentum space to coordinate space as 
\begin{eqnarray}
\left\langle P_{1}^{\prime}\right|A_{\sigma}^{\dagger}(p_{1}^{\prime})A_{\mu}(p_{1})\left|P_{1}\right\rangle  & \rightarrow & \left\langle P_{1}^{\prime}\right|A_{\sigma}^{\dagger}(x_{1}^{\prime})A_{\mu}(x_{1})\left|P_{1}\right\rangle ,\nonumber \\
\left\langle P_{2}^{\prime}\right|A_{\rho}^{\dagger}(p_{2}^{\prime})A_{\nu}(p_{2})\left|P_{2}\right\rangle  & \rightarrow & \left\langle P_{2}^{\prime}\right|A_{\rho}^{\dagger}(x_{2}^{\prime})A_{\nu}(x_{2})\left|P_{2}\right\rangle ,
\end{eqnarray}
and changed all coordinate variables to $X_{i}=(x_{i}+x_{i}^{\prime})/2$
and $y_{i}=x_{i}-x_{i}^{\prime}$ for $i=1,2$, then we carried out
the integration over $X_{i}$ after shifting $x_{1}^{\prime}$ and
$x_{2}^{\prime}$ for the photon fields in $\left\langle P_{1}^{\prime}\right|A_{\sigma}^{\dagger}(x_{1}^{\prime})A_{\mu}(x_{1})\left|P_{1}\right\rangle $
and $\left\langle P_{2}^{\prime}\right|A_{\rho}^{\dagger}(x_{2}^{\prime})A_{\nu}(x_{2})\left|P_{2}\right\rangle $
respectively by making use of the formula $A^{\mu}(x+y)=e^{i\hat{p}\cdot x}A^{\mu}(y)e^{-i\hat{p}\cdot x}$.

Inserting (\ref{eq:photon-me-x-space}) into Eq. (\ref{eq:cross-section-2}),
the cross section is rewritten as 
\begin{eqnarray}
\sigma & = & \frac{1}{8(2\pi)^{8}}\int d^{2}\mathbf{b}_{T}d^{2}\mathbf{b}_{1T}d^{2}\mathbf{b}_{2T}\int\frac{d^{3}k_{1}}{(2\pi)^{3}2E_{k1}}\frac{d^{3}k_{2}}{(2\pi)^{3}2E_{k2}}\nonumber \\
 &  & \times\int d^{2}\mathbf{P}_{1T}d^{2}\mathbf{P}_{2T}d^{2}\mathbf{P}_{1T}^{\prime}d^{2}\mathbf{P}_{2T}^{\prime}\frac{1}{v\sqrt{E_{P1}E_{P2}E_{P1^{\prime}}E_{P2^{\prime}}}}\nonumber \\
 &  & \times\phi_{T}(\mathbf{P}_{1T})\phi_{T}(\mathbf{P}_{2T})\phi_{T}^{*}(\mathbf{P}_{1T}^{\prime})\phi_{T}^{*}(\mathbf{P}_{2T}^{\prime})G^{2}\left[(P_{1}^{\prime z}-P_{A1}^{z})^{2}\right]\nonumber \\
 &  & \times e^{-i\mathbf{b}_{1T}\cdot\boldsymbol{\Delta}_{1T}}e^{-i\mathbf{b}_{2T}\cdot\boldsymbol{\Delta}_{2T}}\delta^{(2)}\left(\mathbf{b}_{T}-\mathbf{b}_{1T}+\mathbf{b}_{2T}\right)\nonumber \\
 &  & \times\int\frac{d^{4}p_{1}}{(2\pi)^{4}}\frac{d^{4}p_{2}}{(2\pi)^{4}}\int d^{4}y_{1}d^{4}y_{2}\exp\left(ip_{1}\cdot y_{1}+ip_{2}\cdot y_{2}\right)\nonumber \\
 &  & \times\left\langle P_{1}^{\prime}\right|A_{\sigma}^{\dagger}(0)A_{\mu}\left(y_{1}\right)\left|P_{1}\right\rangle \left\langle P_{2}^{\prime}\right|A_{\rho}^{\dagger}(0)A_{\nu}\left(y_{2}\right)\left|P_{2}\right\rangle \nonumber \\
 &  & \times L^{\mu\nu;\sigma\rho}\left(p_{1},p_{2};p_{1}-P_{1}+P_{1}^{\prime},p_{2}-P_{2}+P_{2}^{\prime};k_{1},k_{2}\right)\nonumber \\
 &  & \times(2\pi)^{4}\delta^{(4)}\left(p_{1}+p_{2}-k_{1}-k_{2}\right).\label{eq:cross-section-3}
\end{eqnarray}
Equations (\ref{eq:cross-section-2},\ref{eq:cross-section-3}) are
our starting point.


\section{Some useful formula in light-cone coordinates \label{sec:Lightcone}}

A four-vector or a four-tensor can be written in light-cone variables.
For example, the coordinate is written as $x^{\mu}=(x^{+},x^{-},\mathbf{x}_{T})$
with $x^{\pm}=(x^{0}\pm x^{3})/\sqrt{2}$ and $\mathbf{x}_{T}=(x^{1},x^{2})$.
The Minkowski metric tensor becomes

\begin{equation}
g_{\mu\nu}=\left(\begin{array}{cccc}
0 & 1 & 0 & 0\\
1 & 0 & 0 & 0\\
0 & 0 & -1 & 0\\
0 & 0 & 0 & -1
\end{array}\right),
\end{equation}
with $\mu=+,-,1,2$. So we have $x_{\mu}=g_{\mu\nu}x^{\nu}=(x^{-},x^{+},-\mathbf{x}_{T})$.
It is convenient to decompose an arbitrary four-vector $a^{\mu}$
by light-like Sudakov vectors $n_{+}^{\mu}$ and $n_{-}^{\mu}$, satisfying
$n_{+}^{2}=n_{-}^{2}=0$ and $n_{+}\cdot n_{-}=1$,
\begin{equation}
a^{\mu}=a^{+}n_{+}^{\mu}+a^{-}n_{-}^{\mu}+a_{T}^{\mu},
\end{equation}
where $a^{\pm}=(a^{0}\pm a^{3})/\sqrt{2}$ and $a_{T}^{\mu}=g_{T}^{\mu\nu}a_{\nu}$
with the space-like transverse projector

\begin{equation}
g_{T}^{\mu\nu}=g^{\mu\nu}-n_{+}^{\mu}n_{-}^{\nu}-n_{+}^{\nu}n_{-}^{\mu}.
\end{equation}
The inner product of two vectors is $a\cdot b=a^{-}b^{+}+a^{+}b^{-}-\boldsymbol{a}_{T}\cdot\boldsymbol{b}_{T}$.
The explicit forms of light-like Sudakov vectors are 
\begin{eqnarray}
n_{+}^{\mu} & = & (1,0,0,0)_{\mathrm{light}\:\mathrm{cone}}=\frac{1}{\sqrt{2}}(1,0,0,1)_{\mathrm{normal}\:\mathrm{coordinate}},\nonumber \\
n_{-}^{\mu} & = & (0,1,0,0)_{\mathrm{light}\:\mathrm{cone}}=\frac{1}{\sqrt{2}}(1,0,0,-1)_{\mathrm{normal}\:\mathrm{coordinate}},\label{eq:sudakov-vector}
\end{eqnarray}

The four-momenta of colliding nuclei can be written as 
\begin{equation}
P_{A1,A2}^{\mu}=M_{1,2}u_{1,2}^{\mu},\label{eq:def_P1_M}
\end{equation}
where $M_{1,2}$ and $u_{1,2}^{\mu}$ are masses and four-velocities
of two nuclei respectively and $u_{1,2}^{\mu}$ are given by 
\begin{eqnarray}
u_{1}^{\mu} & = & \gamma_{1}\left(1,0,0,v_{1}\right),\nonumber \\
u_{2}^{\mu} & = & \gamma_{2}\left(1,0,0,-v_{2}\right),\label{eq:velocities}
\end{eqnarray}
with $\gamma_{1,2}=1/\sqrt{1-v_{1,2}^{2}}$ being the Lorentz factor.
Then the four-momenta of two nuclei can be written in the form
\begin{eqnarray}
P_{A1}^{\mu} & = & P_{A1}^{+}n_{+}^{\mu}+\frac{M_{1}^{2}}{2P_{A1}^{+}}n_{-}^{\mu},\nonumber \\
P_{A2}^{\mu} & = & \frac{M_{2}^{2}}{2P_{A2}^{-}}n_{+}^{\mu}+P_{A2}^{-}n_{-}^{\mu},
\end{eqnarray}
where light-cone variables are 
\begin{equation}
P_{A1}^{+}=\frac{M_{1}\gamma_{1}}{\sqrt{2}}(1+v_{1}),\;\;P_{A2}^{-}=\frac{M_{2}\gamma_{2}}{\sqrt{2}}(1+v_{2}).
\end{equation}
For high energy nuclei with $v_{1,2}\sim1$, we have $P_{A1}^{+}\gg M_{1}$
and $P_{A2}^{-}\gg M_{2}$.

From the classical photon field (\ref{eq:classical_01}), we have
$p_{i}\cdot u_{i}=p_{i}^{\prime}\cdot u_{i}=0$ for $i=1,2$. Then
$p_{1}$ and $p_{2}$ can be decomposed as
\begin{eqnarray}
p_{1}^{\mu} & = & p_{1}^{+}n_{+}^{\mu}+p_{1}^{-}n_{-}^{\mu}+p_{1T}^{\mu},\nonumber \\
p_{2}^{\mu} & = & p_{2}^{+}n_{+}^{\mu}+p_{2}^{-}n_{-}^{\mu}+p_{2T}^{\mu},
\end{eqnarray}
where 
\begin{equation}
p_{1}^{\pm}=\frac{\omega_{1}}{\sqrt{2}}\left(1\pm\frac{1}{v_{1}}\right),\;\;p_{2}^{\pm}=\frac{\omega_{2}}{\sqrt{2}}\left(1\mp\frac{1}{v_{2}}\right).\label{eq:p1_2_light_cone_01}
\end{equation}
At high energies, we have $v_{1,2}\rightarrow1$, so it is easy to
verify 
\begin{equation}
\frac{p_{1}^{+}}{\omega_{1}},\frac{p_{2}^{-}}{\omega_{2}}\sim\mathcal{O}(1),\;\;\frac{p_{1}^{-}}{\omega_{1}},\frac{p_{2}^{+}}{\omega_{2}}\sim\mathcal{O}(\gamma^{-2}).\label{eq:p+p--energy}
\end{equation}
On the other hand, in the rest frame of a nucleus, it is assumed 
\begin{equation}
p_{i}^{2}\sim\mathcal{O}(x_{i}^{2}M_{i}^{2}),
\end{equation}
therefore, we have 

\begin{equation}
\frac{\mathbf{p}_{iT}}{\omega_{i}}\sim\frac{x_{i}M_{i}}{\omega_{i}}\sim\mathcal{O}(\gamma^{-1}),\label{eq:pt-energy}
\end{equation}
which means photons are almost on-shell.

We assume a collision of two identical nuclei, so we have $v_{1}=v_{2}$.
To simplify $u_{1\mu}u_{2\nu}L^{\mu\nu}$ in Eq. (\ref{eq:cross-section-photon-impact}),
we use Eq. (\ref{eq:relation_Gamma_01}) and obtain 
\begin{eqnarray}
L^{0\nu}(p_{1},p_{2},k_{1},k_{2}) & = & \frac{p_{1}^{i}L^{i\nu}+p_{1}^{z}L^{z\nu}}{\omega_{1}},\nonumber \\
L^{\mu0}(p_{1},p_{2},k_{1},k_{2}) & = & \frac{p_{2}^{i}L^{\mu i}+p_{2}^{z}L^{\mu z}}{\omega_{2}},
\end{eqnarray}
where $i=x,y$ denotes two transverse directions. From Eq. (\ref{eq:p1_2_light_cone_01})
or $p_{1,2}^{z}/\omega_{1,2}=\pm1/v$ (the upper/lower sign corresponds
to nucleus 1 and 2 respectively), $u_{1\mu}u_{2\nu}L^{\mu\nu}$ can
be rewritten as 
\begin{eqnarray}
u_{1\mu}u_{2\nu}L^{\mu\nu} & = & \gamma^{2}\frac{p_{1}^{i}}{\omega_{1}}\frac{p_{2}^{j}}{\omega_{2}}L^{ij}-\frac{1}{v}\left(\frac{p_{1}^{i}}{\omega_{1}}L^{iz}-\frac{p_{2}^{i}}{\omega_{2}}L^{zi}\right)-\frac{1}{\gamma^{2}v^{2}}L^{zz},\label{eq:uuGamma_01}
\end{eqnarray}
where $i,j=x,y$ denote two transverse directions. The result of (\ref{eq:uuGamma_01})
is consistent with Ref. \citep{Vidovic:1992ik}. %
{} We can also use the light-cone coordinate and obtain 
\begin{eqnarray}
L^{-\nu}(p_{1},p_{2},k_{1},k_{2}) & = & -\frac{p_{1}^{-}L^{+\nu}-p_{1}^{i}L^{i\nu}}{p_{1}^{+}},\nonumber \\
L^{\mu+}(p_{1},p_{2},k_{1},k_{2}) & = & -\frac{p_{2}^{+}L^{\mu-}-p_{2}^{j}L^{\mu j}}{p_{2}^{-}}.
\end{eqnarray}
So we evaluate $u_{1\mu}u_{2\nu}L^{\mu\nu}$ as 
\begin{eqnarray}
u_{1\mu}u_{2\nu}L^{\mu\nu} & = & \gamma^{2}v^{2}\frac{p_{1}^{i}}{\omega_{1}}\frac{p_{2}^{j}}{\omega_{2}}L^{ij}\nonumber \\
 &  & -2\gamma^{2}v^{2}\left(\frac{p_{1}^{i}}{\omega_{1}}\frac{p_{2}^{+}}{\omega_{2}}L^{i-}+\frac{p_{1}^{-}}{\omega_{1}}\frac{p_{2}^{j}}{\omega_{2}}L^{+j}\right)\nonumber \\
 &  & +4\gamma^{2}v^{2}\frac{p_{1}^{-}}{\omega_{1}}\frac{p_{2}^{+}}{\omega_{2}}L^{+-},\label{eq:uuGamma_02}
\end{eqnarray}
It seems that there is an extra factor $v^{2}$ in the first term
of Eq. (\ref{eq:uuGamma_02}) in comparison with Eq. (\ref{eq:uuGamma_01}),
it is straightforward to prove that Eq. (\ref{eq:uuGamma_02}) and
(\ref{eq:uuGamma_01}) are equivalent. According to Eqs. (\ref{eq:p+p--energy},\ref{eq:pt-energy}),
at the relativistic limit $v\rightarrow1$ or $\gamma\rightarrow\infty$,
the first, second and third term of (\ref{eq:uuGamma_02}) are $O(1)$,
$O(\gamma^{-1})$ and $O(\gamma^{-2})$ respectively. Therefore the
leading order contribution comes from the first term.

\section{TMD Correlation functions in classical field approximation \label{sec:Correlation-functions-in}}

The gauge invariant correlation function for an unpolarized nucleus
is defined as 
\begin{equation}
\mathcal{W}^{\mu\nu;\rho\sigma}(P,p)=\int\frac{d^{4}\xi}{(2\pi)^{4}}e^{ip\cdot\xi}\left\langle P\right|F^{\mu\nu}(0)F^{\rho\sigma}(\xi)\left|P\right\rangle ,\label{eq:def_corrlation_fun_01}
\end{equation}
where $F^{\mu\nu}=\partial^{\mu}A^{\nu}-\partial^{\nu}A^{\mu}$ is
the field strength tensor of the EM field, and $P=(E_{P},0,0,P^{z})$
is the four-momentum of a nucleus moving in $+z$ direction. The requirement
that $\mathcal{W}^{\mu\nu;\rho\sigma}(P,p)$ must be Hermitian and
parity-even leads to the following parametrization 
\begin{eqnarray}
\mathcal{W}^{\mu\nu;\rho\sigma}(P,p) & = & X_{1}\epsilon^{\mu\nu\alpha\beta}\epsilon_{\alpha\beta}^{\quad\rho\sigma}+\frac{X_{2}}{M^{2}}P^{[\mu}g^{\nu][\rho}P^{\sigma]}+\frac{X_{3}}{M^{2}}p^{[\mu}g^{\nu][\rho}p^{\sigma]}\nonumber \\
 &  & +\frac{X_{4}+iX_{5}}{M^{2}}P^{[\mu}g^{\nu][\rho}p^{\sigma]}+\frac{X_{4}-iX_{5}}{M^{2}}p^{[\mu}g^{\nu][\rho}P^{\sigma]}\nonumber \\
 &  & +\frac{X_{6}}{M^{4}}P^{[\mu}p^{\nu]}P^{[\rho}p^{\sigma]},
\end{eqnarray}
where $A^{[\mu}B^{\nu]}=A^{\mu}B^{\nu}-B^{\mu}A^{\nu}$ and all coefficients
$X_{i}$ are real functions of $P$ and $p$. We work in light-cone
coordinate system and focus on $\mathcal{W}^{+\mu;+\nu}(P,p)$ which
is related to $S_{\sigma\mu}(P,p)$ in (\ref{eq:TMD_unpol_01}) in
the light-cone gauge 
\begin{eqnarray}
\mathcal{W}^{+\mu;+\nu}(P,p) & = & -g_{T}^{\mu\nu}\left(\frac{P^{+}}{M}\right)^{2}\left(X_{2}+2xX_{4}+x^{2}X_{3}\right)+\left(\frac{P^{+}}{M}\right)^{2}\frac{p_{T}^{\mu}p_{T}^{\nu}}{M^{2}}X_{6}\nonumber \\
 &  & +\frac{P^{+}}{M}\frac{p_{T}^{\mu}n_{-}^{\nu}+p_{T}^{\nu}n_{-}^{\mu}}{M}\left[X_{4}+xX_{3}+\left(\frac{\sigma}{2}-x\right)X_{6}\right]-\frac{P^{+}}{M}\frac{p_{T}^{\text{[}\mu}n_{-}^{\nu]}}{M}iX_{5}\nonumber \\
 &  & +n_{-}^{\mu}n_{-}^{\nu}\left[2X_{1}+X_{2}+xX_{4}+x(X_{4}+xX_{3})\right.\nonumber \\
 &  & \left.+2\left(\frac{\sigma}{2}-x\right)(X_{4}+xX_{3})+\left(\frac{\sigma}{2}-x\right)^{2}X_{6}\right],\label{eq:w+mu+nu}
\end{eqnarray}
where we have used $\sigma=2(P\cdot p)/M^{2}$ and $x=p^{+}/P^{+}$.

The twist-2 correlation function comes from transverse components
$\mathcal{W}^{+i;+j}$ 
\begin{equation}
\mathcal{W}^{+i;+j}=-g_{T}^{ij}\left(\frac{P^{+}}{M}\right)^{2}\left(X_{2}+2xX_{4}+x^{2}X_{3}\right)+\left(\frac{P^{+}}{M}\right)^{2}\frac{p_{T}^{i}p_{T}^{j}}{M^{2}}X_{6}.\label{eq:twist2_02}
\end{equation}
Integrating over $p^{-}$ for $\mathcal{W}^{+i;+j}$, we have the
following parametrization 
\begin{eqnarray}
 &  & \int dp^{-}\mathcal{W}^{+i;+j}(P,p)\nonumber \\
 & = & \frac{1}{2}P^{+}\left[-g_{T}^{ij}xf^{\gamma}(x,\mathbf{p}_{T}^{2})+\left(\frac{p_{T}^{i}p_{T}^{j}}{M^{2}}+\frac{\mathbf{p}_{T}^{2}}{2M^{2}}g_{T}^{ij}\right)xh_{T}^{\gamma}(x,\mathbf{p}_{T}^{2})\right],\label{eq:f-ht-w}
\end{eqnarray}
where $f^{\gamma}(x,\mathbf{p}_{T}^{2})$ and $h_{T}^{\gamma}(x,\mathbf{p}_{T}^{2})$
are distributions defined as 
\begin{eqnarray}
xf^{\gamma}(x,\mathbf{p}_{T}^{2}) & = & \frac{2P^{+}}{M^{2}}\int dp^{-}\left(X_{2}+2xX_{4}+x^{2}X_{3}+\frac{\mathbf{p}_{T}^{2}}{2M^{2}}X_{6}\right),\nonumber \\
xh_{T}^{\gamma}(x,\mathbf{p}_{T}^{2}) & = & \frac{2P^{+}}{M^{2}}\int dp^{-}X_{6}.\label{eq:f_h_expression_01}
\end{eqnarray}
The twist-3 contribution of $\mathcal{W}^{+\mu;+\nu}$ comes from
following components 
\begin{eqnarray}
\mathcal{W}^{+i;+-} & = & \frac{P^{+}}{M}\frac{p_{T}^{i}}{M}\left[X_{4}+xX_{3}+\left(\frac{\sigma}{2}-x\right)X_{6}-iX_{5}\right]\nonumber \\
\mathcal{W}^{ij;k+} & = & \frac{-g_{T}^{k[i}p_{T}^{j]}}{M}\frac{P^{+}}{M}(X_{4}+xX_{3}-iX_{5}),
\end{eqnarray}
The twist-$4$ contribution reads 
\begin{eqnarray}
\mathcal{W}^{+-;+-} & = & 2X_{1}+X_{2}+xX_{4}+x(X_{4}+xX_{3})\nonumber \\
 &  & +2\left(\frac{\sigma}{2}-x\right)(X_{4}+xX_{3})+\left(\frac{\sigma}{2}-x\right)^{2}X_{6}.
\end{eqnarray}
More details and discussions can be found in Ref. \citep{Mulders:2000sh}
and references therein.

Now we take the classical field approximation. We transform Eq. (\ref{eq:w+mu+nu})
to momentum space and obtain 
\begin{eqnarray}
\mathcal{W}^{+\mu;+\nu}(P,p) & = & \frac{1}{(2\pi)^{4}}\int\frac{d^{4}p^{\prime}}{(2\pi)^{4}}\left\langle P\right|F^{+\mu}(p^{\prime})F^{+\nu}(p)\left|P\right\rangle \nonumber \\
 & = & \frac{1}{(2\pi)^{4}}\int\frac{d^{4}p^{\prime}}{(2\pi)^{4}}\left\langle P\right|\left[p^{\prime+}A^{\dagger\mu}(p^{\prime})-p^{\prime\mu}A^{\dagger+}(p^{\prime})\right]\nonumber \\
 &  & \times\left[p^{+}A^{\nu}(p)-p^{\nu}A^{+}(p)\right]\left|P\right\rangle \nonumber \\
 & = & \frac{1}{(2\pi)^{4}}\int\frac{d^{4}p^{\prime}}{(2\pi)^{4}}p^{\prime+}p^{+}\left\langle P\right|A^{\dagger\mu}(p^{\prime})A^{\nu}(p)\left|P\right\rangle ,\label{eq:w+mu+nu-1}
\end{eqnarray}
where we have used the light-cone gauge $A^{+}=0$. In the light-cone
gauge the classical field has the form 
\begin{eqnarray}
A^{\mu}(p) & = & 2\pi Ze\delta(p\cdot u)\frac{F(-p^{2})}{-p^{2}}\left[u^{\mu}-\frac{p^{\mu}}{p^{+}}u^{+}\right],
\end{eqnarray}
as the solution to the Maxwell equation. Here $u^{\mu}=\gamma(1,0,0,v)$
is the four-velocity of the nucleus satisfying $P^{\mu}=Mu^{\mu}$.
We use the same ansatz as (\ref{eq:photon-corr}) for the matrix element
in Eq. (\ref{eq:w+mu+nu-1}) 
\begin{eqnarray}
\left\langle P\right|A^{\dagger\mu}(p^{\prime})A^{\nu}(p)\left|P\right\rangle  & \approx & 2MA^{*\mu}(p^{\prime})A^{\nu}(p)\nonumber \\
 &  & \times(2\pi)^{3}\delta\left(p\cdot\overline{u}-p^{\prime}\cdot\overline{u}\right)\delta^{(2)}\left(\mathbf{p}_{T}-\mathbf{p}_{T}^{\prime}\right),
\end{eqnarray}
where $\overline{u}^{\mu}\equiv\gamma(v,0,0,1)$ satisfying $u\cdot\overline{u}=0$.
Using the above formula in (\ref{eq:w+mu+nu-1}) we obtain the transverse
component 
\begin{eqnarray}
\mathcal{W}^{+i;+j}(P,p) & = & \frac{1}{2\pi}\int\frac{d^{4}p^{\prime}}{(2\pi)^{4}}p^{\prime+}p^{+}2MA^{*i}(p^{\prime})A^{j}(p)\delta\left(p\cdot\overline{u}-p^{\prime}\cdot\overline{u}\right)\delta^{(2)}\left(\mathbf{p}_{T}-\mathbf{p}_{T}^{\prime}\right)\nonumber \\
 & = & \frac{1}{2\pi}(2\pi Ze)^{2}2M\int\frac{d^{4}p^{\prime}}{(2\pi)^{4}}p^{\prime+}p^{+}\nonumber \\
 &  & \times\delta(p^{\prime}\cdot u)\delta(p\cdot u)\frac{F^{*}(-p^{\prime2})}{-p^{\prime2}}\frac{F(-p^{2})}{-p^{2}}\left[u^{i}-\frac{p^{\prime i}}{p^{\prime+}}u^{+}\right]\left[u^{j}-\frac{p^{j}}{p^{+}}u^{+}\right]\nonumber \\
 &  & \times\delta\left(p\cdot\overline{u}-p^{\prime}\cdot\overline{u}\right)\delta^{(2)}\left(\mathbf{p}_{T}-\mathbf{p}_{T}^{\prime}\right)\nonumber \\
 & = & \frac{1}{4\pi^{3}}MZ^{2}e^{2}(u^{+})^{2}\delta(p\cdot u)\left|\frac{F(-p^{2})}{-p^{2}}\right|^{2}p_{T}^{i}p_{T}^{j},\label{eq:w+i+j-light-cone}
\end{eqnarray}
where we have used $d^{4}p^{\prime}=d(p^{\prime}\cdot u)d(-p^{\prime}\cdot\overline{u})d^{2}\mathbf{p}_{T}^{\prime}$
in carrying out the integral over $p^{\prime}$, and the constraints
by the delta-functions give $p^{\prime\mu}=p^{\mu}$.

Then we take an integral over $p^{-}$ for $\mathcal{W}^{+i;+j}(P,p)$
and obtain 
\begin{equation}
\int dp^{-}\mathcal{W}^{+i;+j}(P,p)=\frac{1}{4\pi^{3}}MZ^{2}e^{2}u^{+}\left|\frac{F(-p^{2})}{-p^{2}}\right|^{2}p_{T}^{i}p_{T}^{j},
\end{equation}
where $p$ denotes the momentum that satisfies $p\cdot u=0$, in Sect.
\ref{sec:Classical-field-approximation} we denote it as $\overline{p}$
as defined in Eq. (\ref{eq:orth-momenta}) but here we suppressed
the bar to simplify the notation. Comparing with Eq. (\ref{eq:f-ht-w})
we can extract 
\begin{eqnarray}
xh_{T}^{\gamma}(x,\mathbf{p}_{T}^{2}) & = & \frac{2Z^{2}\alpha}{\pi^{2}}M^{2}\left|\frac{F(-p^{2})}{-p^{2}}\right|^{2},\nonumber \\
xf^{\gamma}(x,\mathbf{p}_{T}^{2}) & = & \frac{\mathbf{p}_{T}^{2}}{2M^{2}}xh_{T}^{\gamma}(x,\mathbf{p}_{T}^{2})=\frac{Z^{2}\alpha}{\pi^{2}}\mathbf{p}_{T}^{2}\left|\frac{F(-p^{2})}{-p^{2}}\right|^{2}.
\end{eqnarray}

\section{Independent variables for numerical integration of cross sections
\label{sec:Integral-variables}}

If the wave packets are very narrow in momentum we can make an approximation
in Eq. (\ref{eq:cs-photon-impact}): $G_{T}\left[(\mathbf{p}_{iT}^{\prime}-\mathbf{p}_{iT})^{2}\right]\approx1$
for $i=1,2$. In this case we can carry out the integrals over $\mathbf{b}_{1T}$,
$\mathbf{b}_{2T}$ and $\mathbf{p}_{2T}^{\prime}$ to obtain 
\begin{eqnarray}
\sigma & = & \frac{Z^{4}e^{4}}{2\gamma^{4}v^{3}}\int d^{2}\mathbf{b}_{T}\int\frac{d\omega_{1}d^{2}\mathbf{p}_{1T}}{(2\pi)^{3}}\frac{d\omega_{2}d^{2}\mathbf{p}_{2T}}{(2\pi)^{3}}\frac{d^{2}\mathbf{p}_{1T}^{\prime}}{(2\pi)^{2}}e^{-i\mathbf{b}_{T}\cdot(\mathbf{p}_{1T}^{\prime}-\mathbf{p}_{1T})}\nonumber \\
 &  & \times\frac{F^{*}(-p_{1}^{\prime2})}{-p_{1}^{\prime2}}\frac{F(-p_{1}^{2})}{-p_{1}^{2}}\frac{F^{*}(-p_{2}^{\prime2})}{-p_{2}^{\prime2}}\frac{F(-p_{2}^{2})}{-p_{2}^{2}}\nonumber \\
 &  & \times\int\frac{d^{3}k_{1}}{(2\pi)^{3}2E_{k1}}\frac{d^{3}k_{2}}{(2\pi)^{3}2E_{k2}}\sum_{\textrm{spin of }l,\overline{l}}\left[u_{1\mu}u_{2\nu}L^{\mu\nu}(p_{1},p_{2};k_{1},k_{2})\right]\nonumber \\
 &  & \times\left[u_{1\sigma}u_{2\rho}L^{\sigma\rho*}(p_{1}^{\prime},p_{2}^{\prime};k_{1},k_{2})\right](2\pi)^{4}\delta^{(4)}\left(p_{1}+p_{2}-k_{1}-k_{2}\right),
\end{eqnarray}
where $\mathbf{p}_{2T}^{\prime}=\mathbf{p}_{1T}+\mathbf{p}_{2T}-\mathbf{p}_{1T}^{\prime}$,
$\omega_{1}^{\prime}=\omega_{1}$ and $\omega_{2}^{\prime}=\omega_{2}$.
Note that the photon momenta $p_{1},p_{2},p_{1}^{\prime},p_{2}^{\prime}$
are momenta that satisfy $p_{i}\cdot u=0$ and $p_{i}^{\prime}\cdot u=0$
for $i=1,2$, in Sect. \ref{sec:Classical-field-approximation} we
denote them as $\overline{p}_{1},\overline{p}_{2},\overline{p}_{1}^{\prime},\overline{p}_{2}^{\prime}$
as defined in Eq. (\ref{eq:orth-momenta}) but here we suppressed
the bars to simplify the notation. Then we can complete the integrals
over $\omega_{1}$, $\omega_{2}$ and $\mathbf{p}_{2T}$ to remove
the delta functions, fixing $\omega_{1}$ and $\omega_{2}$ as 
\begin{eqnarray}
\omega_{1} & = & \frac{1}{2}[k_{1}^{0}+k_{2}^{0}+(k_{1}^{z}+k_{2}^{z})v],\nonumber \\
\omega_{2} & = & \frac{1}{2}[k_{1}^{0}+k_{2}^{0}-(k_{1}^{z}+k_{2}^{z})v],\label{eq:omega12_01}
\end{eqnarray}
and fixing $\mathbf{p}_{2T}$ and $\mathbf{p}_{2T}^{\prime}$ as $\mathbf{p}_{2T}=\mathbf{k}_{1T}+\mathbf{k}_{2T}-\mathbf{p}_{1T}$
and $\mathbf{p}_{2T}^{\prime}=\mathbf{k}_{1T}+\mathbf{k}_{2T}-\mathbf{p}_{1T}^{\prime}$.
This gives the cross section of the form 
\begin{eqnarray}
\sigma & = & \frac{Z^{4}\alpha^{2}}{64\pi^{4}\gamma^{4}v^{2}}\int dy_{1}d^{2}\mathbf{k}_{1T}dy_{2}d^{2}\mathbf{k}_{2T}\nonumber \\
 &  & \times\int d^{2}\mathbf{b}_{T}\int\frac{d^{2}\mathbf{p}_{1T}}{(2\pi)^{2}}\frac{d^{2}\mathbf{p}_{1T}^{\prime}}{(2\pi)^{2}}e^{-i\mathbf{b}_{T}\cdot(\mathbf{p}_{1T}^{\prime}-\mathbf{p}_{1T})}\nonumber \\
 &  & \times\frac{F^{*}(-p_{1}^{\prime2})}{-p_{1}^{\prime2}}\frac{F(-p_{1}^{2})}{-p_{1}^{2}}\frac{F^{*}(-p_{2}^{\prime2})}{-p_{2}^{\prime2}}\frac{F(-p_{2}^{2})}{-p_{2}^{2}}\nonumber \\
 &  & \times\sum_{\textrm{spin of }l,\overline{l}}\left[u_{1\mu}u_{2\nu}L^{\mu\nu}(p_{1},p_{2};k_{1},k_{2})\right]\nonumber \\
 &  & \times\left[u_{1\sigma}u_{2\rho}L^{\sigma\rho*}(p_{1}^{\prime},p_{2}^{\prime};k_{1},k_{2})\right],\label{eq:cross-section-var}
\end{eqnarray}
where we have used the rapidity and transverse momentum as independent
variables for lepton momenta so that the lepton momenta $k_{1}$ and
$k_{2}$ can be expressed as 
\begin{equation}
k^{\mu}=\frac{m_{T}}{\sqrt{2}}e^{y}n_{+}^{\mu}+\frac{m_{T}}{\sqrt{2}}e^{-y}n_{-}^{\mu}+k_{T}^{\mu}.
\end{equation}
Here the rapidity and transverse mass are defined as 
\begin{eqnarray}
y & = & \frac{1}{2}\ln\frac{E+k_{z}}{E-k_{z}}=\frac{1}{2}\ln\frac{k^{+}}{k^{-}},\nonumber \\
m_{T}^{2} & = & \mathbf{k}_{T}^{2}+m^{2}=E^{2}-k_{z}^{2}=2k^{+}k^{-}.
\end{eqnarray}
Then we have $d^{3}k/E_{k}=dyd^{2}\mathbf{k}_{T}$.

Now we choose independent integration variables of the cross section.
The impact parameter is written as 
\begin{equation}
\mathbf{b}_{T}=(b_{T}\cos\phi_{b},b_{T}\sin\phi_{b}).
\end{equation}
The transverse momentum shift is defined as 
\begin{equation}
\boldsymbol{\Delta}_{T}\equiv\mathbf{p}_{1T}^{\prime}-\mathbf{p}_{1T}=(\Delta_{T}\cos\phi_{\Delta},\Delta_{T}\sin\phi_{\Delta}),\label{eq:Delta_phi_d_01}
\end{equation}
so the integral over $\mathbf{p}_{1T}^{\prime}$ can be replaced by
$\boldsymbol{\Delta}_{T}$. Since the invariant mass and transverse
momentum spectra of lepton pairs were measured in experiments, we
should define the four-momentum of the lepton pair
\begin{eqnarray}
P_{ee}^{\mu} & = & k_{1}^{\mu}+k_{2}^{\mu}=(P_{ee}^{0},\mathbf{P}_{ee}^{T},P_{ee}^{z})\nonumber \\
 & = & \left(\sqrt{M_{ee}^{2}+(P_{ee}^{T})^{2}}\cosh Y_{ee},P_{ee}^{T}\cos\phi_{ee},P_{ee}^{T}\sin\phi_{ee},\right.\nonumber \\
 &  & \left.\sqrt{M_{ee}^{2}+(P_{ee}^{T})^{2}}\sinh Y_{ee}\right),\label{eq:q_M_ee}
\end{eqnarray}
where $Y_{ee}$ is the rapidity of the lepton pair, and $M_{ee}$
is the invariant mass of the lepton pair given by 
\begin{equation}
M_{ee}^{2}=P_{ee}^{2}=2m_{e}^{2}+2k_{1}\cdot k_{2}=2P_{ee}\cdot k_{2}.\label{eq:mee-pee}
\end{equation}
Instead of $k_{1}^{\mu}$ and $k_{2}^{\mu}$, we choose $P_{ee}^{\mu}$
and two variables of $k_{2}^{\mu}$ as independent variables since
$k_{2}^{\mu}$ satisfies the constraint (\ref{eq:mee-pee}). Two variables
of $k_{2}^{\mu}$ are chosen to be the pseudo-rapidity $\eta_{k2}$
and the azimuthal angle $\phi_{k2}$ of the transverse momentum, so
$k_{2}^{\mu}$ can be expressed by 
\begin{equation}
k_{2}^{\mu}=\left(\sqrt{m_{e}^{2}+k_{2T}^{2}\cosh^{2}\eta_{k2}},k_{2T}\cos\phi_{k2},k_{2T}\sin\phi_{k2},k_{2T}\sinh\eta_{k2}\right).
\end{equation}
Here $k_{2T}$ is obtained by solving Eq. (\ref{eq:mee-pee}) 
\begin{equation}
k_{2T}=\frac{f_{1}M_{ee}^{2}+\sqrt{f_{1}^{2}M_{ee}^{4}-f_{2}f_{3}}}{f_{2}},
\end{equation}
where 
\begin{eqnarray}
f_{1} & = & 2\left(P_{ee}^{T}\cos\phi_{ee}\cos\phi_{k2}+q_{T}\sin\phi_{ee}\sin\phi_{k2}\right.\nonumber \\
 &  & \left.+\sqrt{M_{ee}^{2}+(P_{ee}^{T})^{2}}\sinh Y_{ee}\sinh\eta_{k2}\right),\nonumber \\
f_{2} & = & 4\left[M_{ee}^{2}+(P_{ee}^{T})^{2}\right]\cosh^{2}Y_{ee}\cosh^{2}\eta_{k2}-f_{1}^{2},\nonumber \\
f_{3} & = & -M_{ee}^{4}+4\left[M_{ee}^{2}+(P_{ee}^{T})^{2}\right]m_{e}^{2}\cosh^{2}Y_{ee}.
\end{eqnarray}
One can verfify 
\begin{equation}
dy_{1}d^{2}k_{1T}dy_{2}d^{2}k_{2T}=\mathcal{J}dY_{ee}dP_{ee}^{T}d\phi_{ee}dM_{ee}d\eta_{k2}d\phi_{k2},
\end{equation}
where $\mathcal{J}$ is a function of $P_{ee}^{\mu}$ and $k_{2}^{\mu}$.
From $P_{ee}^{\mu}$ and $k_{2}^{\mu}$, we obtain $k_{1}^{\mu}=P_{ee}^{\mu}-k_{2}^{\mu}$.
The photon momenta are given by 
\begin{eqnarray}
p_{1}^{\mu} & = & \left(\frac{P_{ee}^{0}+vP_{ee}^{z}}{2},p_{1T}\cos\phi_{p1},p_{1T}\sin\phi_{p1},\frac{P_{ee}^{0}+vP_{ee}^{z}}{2v}\right),\nonumber \\
p_{2}^{\mu} & = & \left(\frac{P_{ee}^{0}-vP_{ee}^{z}}{2},P_{ee}^{x}-p_{1T}\cos\phi_{p1},P_{ee}^{y}-p_{1T}\sin\phi_{p1},-\frac{P_{ee}^{0}-vP_{ee}^{z}}{2v}\right),\nonumber \\
p_{1}^{\prime\mu} & = & \left(\frac{P_{ee}^{0}+vP_{ee}^{z}}{2},p_{1T}\cos\phi_{p1}+\Delta_{T}\cos\phi_{\Delta},\right.\nonumber \\
 &  & \left.p_{1T}\sin\phi_{p1}+\Delta_{T}\sin\phi_{\Delta},\frac{P_{ee}^{0}+vP_{ee}^{z}}{2v}\right),\nonumber \\
p_{2}^{\prime\mu} & = & \left(\frac{P_{ee}^{0}-vP_{ee}^{z}}{2},P_{ee}^{x}-p_{1T}\cos\phi_{p1}-\Delta_{T}\cos\phi_{\Delta},\right.\nonumber \\
 &  & \left.P_{ee}^{y}-p_{1T}\sin\phi_{p1}-\Delta_{T}\sin\phi_{\Delta},-\frac{P_{ee}^{0}-vP_{ee}^{z}}{2v}\right),
\end{eqnarray}
where we can choose $p_{1T}$ and $\phi_{p1}$ as independent variables. 

In summary, we can choose following integration variables 
\begin{equation}
(Y_{ee},P_{ee}^{T},\phi_{ee},M_{ee}),(\eta_{k2},\phi_{k2}),(p_{1T},\phi_{p1}),(\Delta_{T},\phi_{\Delta}),(b_{T},\phi_{b}),
\end{equation}
where we used the parenthesis to enclose variables from the same source.
In terms of these independent variables, the cross section (\ref{eq:cross-section-var})
can be rewritten in the form of Eq. (\ref{eq:cross-section-indep-var}).


\bibliographystyle{h-physrev}
\bibliography{UPCref2}

\end{document}